\documentclass[]{raa}            

\usepackage{graphicx,times}             
\usepackage{subfig}
\usepackage{dcolumn}
\usepackage{bm}
\usepackage{float}
\usepackage{xcolor}
\usepackage{tabularx}
\usepackage{amsmath}
\usepackage{amssymb}
\usepackage{natbib}
\usepackage{hyperref}
\begin{document}

   \title{Statefinder diagnostic and constraints on the Palatini $f(R)$ gravity theories}


   \author{Shu-Lei Cao
      \inst{1}
   \and Song Li
      \inst{2}
   \and Hao-Ran Yu
      \inst{3}
   \and Tong-Jie Zhang
      \inst{1,4}
   }

   \institute{Department of Astronomy, Beijing Normal University, Beijing, 100875, China; {\it tjzhang@bnu.edu.cn}\\
        \and
             Department of Physics, Capital Normal University, Beijing 100048, China\\
        \and
             Tsung-Dao Lee Institute, Shanghai Jiao Tong University, 800 Dongchuan Road, Shanghai, 200240, China\\
        \and
             School of Information Management, School of Physics and Electric Information, Shandong Provincial Key Laboratory of Biophysics, Dezhou University, Dezhou 253023,China\\
   }

   \date{accepted~~2017~~Nov. 14}

\abstract{ In this paper, we focus on a series of $f(R)$ gravity theories in Palatini formalism to investigate the probabilities of producing the late-time acceleration for the flat Friedmann-Robertson-Walker (FRW) universe. We apply statefinder diagnostic to these cosmological models for chosen series of parameters to see if they distinguish from one another. The diagnostic involves the statefinder pair $\{r,s\}$, where $r$ is derived from the scale factor $a$ and its higher derivatives with respect to the cosmic time $t$, and $s$ is expressed by $r$ and the deceleration parameter $q$. In conclusion, we find that although two types of $f(R)$ theories: (i) $f(R) = R + \alpha R^m - \beta R^{-n}$ and (ii) $f(R) = R + \alpha \ln R - \beta$ can lead to late-time acceleration, their evolutionary trajectories in the $r-s$ and $r-q$ planes reveal different evolutionary properties, which certainly justify the merits of statefinder diagnostic. Additionally, we utilize the observational Hubble parameter data (OHD) to constrain these models of $f(R)$ gravity. As a result, except for $m=n=1/2$ of (i) case, $\alpha=0$ of (i) case and (ii) case allow $\Lambda$CDM model to exist in 1$\sigma$ confidence region. After adopting statefinder diagnostic to the best-fit models, we find that all the best-fit models are capable of going through deceleration/acceleration transition stage with late-time acceleration epoch, and all these models turn to de-Sitter point ($\{r,s\}=\{1,0\}$) in the future. Also, the evolutionary differences between these models are distinct, especially in $r-s$ plane, which makes the statefinder diagnostic more reliable in discriminating cosmological models.
\keywords{dark energy --- cosmology: miscellaneous --- methods: data analysis}
}

   \authorrunning{S.-L. Cao, S. Li, H.-R. Yu, \& T.-J. Zhang }            
   \titlerunning{Statefinder diagnostic on Palatini $f(R)$ gravity }  

   \maketitle

\section{Introduction}

The observations of type \uppercase\expandafter{\romannumeral1}a supernovae (SN\uppercase\expandafter{\romannumeral1}a) \citep{Perlmutter1999,Riess1998} suggest that the universe is currently at an accelerated expansion epoch that is attributed to the dominant component of the universe, dark energy, which not only has a large negative pressure, but also does not cluster as ordinary matters do. In fact, there is no justification for assuming that dark energy resembles known forms of matter or energy, since it has not been detected directly. Up until now, the physical origin of dark energy as well as its nature remains enigmatic.

The simplest model of dark energy is the cosmological constant $\Lambda$ \citep{Sahni2000,Copeland2006}, whose energy density remains constant with time $\rho_{\Lambda} = \Lambda/8\pi G$ (natural units $c = \hbar = 1$ is used throughout the paper) and whose equation of state (defined as the ratio of pressure to energy density) remains $w = 1$ as the universe evolves. Unfortunately, the model is burdened with the well known cosmological constant problems, namely the fine-tuning problem: why is the energy of the vacuum so much smaller than we estimate it should be? and the cosmic coincidence problem: why is the dark energy density approximately equal to the matter density today? These problems have led many researchers to try different approaches for the dark energy issue. Furthermore, the recent analysis of the SN\uppercase\expandafter{\romannumeral1}a data indicates that the time dependent dark energy gives better fit than the cosmological constant. Instead of assuming the equation of state $w$ is a constant, some authors investigate the dynamical scenarios of dark energy. The most popular model among them is dubbed quintessence \citep{Ratra1988,Peebles1988,Ostriker1995}, which invokes an evolving scalar field $\phi$ with a self-interaction potential $V(\phi)$ minimally coupled to gravity. Recently, Zhao et. al. \citep{Zhao2017} find that the dynamical dark energy model preferred at a 3.5$\sigma$ significance level based on the latest observations. Besides, other scalar-field dark energy models have been studied, including phantom \citep{Caldwell2003,Singh2003}, tachyon \citep{Sen2002,Padmanabhan2002}, quintom \citep{Guo2005,Feng2005}, ghost condensates \citep{Arkani-Hamed2004,Piazza2004}. Also, there are other candidates, for example, Chaplygin gas which attempt to unify dark energy and dark matter \citep{Bento2002,Bento2004}, braneworld models which interpret the acceleration through the fact that the general relativity is formulated in five dimensions instead of the usual four \citep{Csaki2000}, backreaction models that consider dark energy as a backreaction effect of inhomogeneities on the average expansion of the universe \citep{Buchert2000,Rasanen2004,Kolb2006}, and so forth.

On the other hand, more and more researchers have made a great deal of effort to consider modifying Einstein¡¯s general relativity (GR) in order to interpret accelerated expansion of the universe without the existence of dark energy. As is well known, there are numerous ways to generalize Einstein¡¯s theory, in which the most famous alternative to GR is scalar-tensor theory \citep{Brans1961,Wagoner1970}. There are still various proposals, for example, Dvali-Gabadadze-Porrati (DGP) gravity \citep{Dvali2000,Deffayet2002}, $f(R)$ gravity \citep{Kerner1982,Allemandi2004}. The so-called $f(R)$ gravity is a straightforward generalization of the Einstein-Hilbert action by including nonlinear terms in the scalar curvature. It has been shown that some of these additional terms can give accelerating expansion without dark energy \citep{Carroll2004}.

Generally, in deriving the Einstein field equations there are two different variational principles that one can apply to the Einstein-Hilbert action, viz., the metric and the Palatini approach. The choice of the variational principle is usually referred to as a formalism, so one can use the metric formalism and the Palatini formalism. In the metric formalism, the connection is assumed to be the Christoffel symbol defined in terms of the metric and the action is only varied with respect to the metric. While in the latter, the metric and the connection are both treated as independent variables, one varies the action with respect to both of them. In fact, for an action which is linear with the Ricci scalar $R$, both approaches are equivalent, and the theory reduces to GR. However, when the action includes nonlinear functions of $R$, different field equations are derived from the two methods.

It was pointed out by Dolgov and Kawasaki that the fourth order equations in the metric formalism suffer serious instability problem \citep{Dolgov2003,Soussa2004,Woodard2007}, however, the Palatini formalism provides second order field equations, which are free from the instability problem mentioned above \citep{Meng2003,Meng2004}. Additionally, for the metric approach, the models of the type $f(R) = R-\beta/R^n$ are incompatible with the solar system experiments \citep{Chiba2003} and have the correct Newtonian limit seemed to be a controversial issue \citep{Sotiriou2006,Sotiriou2006G}. Another important point is that these models can not produce a standard matter-dominated era followed by an accelerating expansion \citep{Amendola2007,Amendola2007P}. While, for the Palatini approach the models satisfy the solar system tests but also have the correct Newtonian limit \citep{Sotiriou2006G}. Furthermore, it has been shown that the above type can produce the sequence of radiation-dominated, matter-dominated and late accelerating phases in \citep{Fay2007}. Thus, as already mentioned, the Palatini approach seems appealing though some issues are controversial, such as the instability problems \citep{Sotiriou2006G,Cembranos2006}. Anyhow, we concentrate on the Palatini formalism.

In addition, since more and more cosmological models have been proposed, the problem of discriminating different models is emergent. In order to tackle this issue, a sensitive and robust diagnosis for dark energy
models is required. It is well known that the equation of state $w$ is able to discriminate some of the dark energy models, for example, the cosmological constant $\Lambda$ with $w=-1$, the quintessence with $w>-1$, the phantom with $w<-1$, and so on. However, for some geometrical models arising from modifications to the gravitational part of Einstein¡¯s theory, the equation of state $w$ no longer plays the essential role and its ambit becomes ambiguous. Therefore, a new diagnosis is requisite to distinguish all classes of cosmological models. In order to achieve this goal, Sahni et al. \citep{Sahni2003} introduce the statefinder pair $\{r,s\}$, where $r$ is derived from the scale factor $a$ and its higher derivatives with respect to the cosmic time $t$, and $s$ is expressed by $r$ and the deceleration parameter $q \equiv -a\ddot{a}/\dot{a}^2$. Thus, the statefinder is a ``geometrical'' diagnostic in the sense that it depends upon the scale factor and hence upon the metric describing spacetime. Based on different cosmological models, distinctions of the evolutionary trajectories in the $r-s$ plane are vivid, which means that the statefinder diagnostic is possibly valid for discriminating different cosmological models. In recent works \citep{Alam2003,Zhang2005,Setare2007,Yi2007}, the statefinder diagnostic has been successfully demonstrated that it can differentiate a series of cosmological models, including the cosmological constant, the quintessence, the phantom, the Chaplygin gas, the holographic dark energy models, the interacting dark energy models, and so forth.

In this paper, we focus on a flat Friedmann-Robertson-Walker (FRW) universe of the $f(R)$ theory in Palatini formalism and consider a number of $f(R)$ theories recently proposed in the literature. In the meantime, we apply the statefinder diagnostic to these $f(R)$ theories. Two types of $f(R)$ theories: (i) $f(R) = R + \alpha R^m - \beta R^{-n}$ and (ii) $f(R) = R + \alpha \ln R - \beta$ are taken into account. Consequently, we find that the models in the Palatini $f(R)$ gravity can be distinguished from one another, as well as $\Lambda$CDM model. In addition, we employ the observational Hubble parameter data (OHD), which are obtained by the differential galactic ages method and the radial Baryon Acoustic Oscillation (BAO) method, to make a combinational constraint. Thereafter, with the best-fit results, we procure the evolutionary trajectories through statefinder diagnostic. Eventually, the results indicate that not only can they demonstrate the possibilities of the late-time acceleration , but also they can demonstrate the limpid distinctions between models.

This paper is organized as follows: In Section \ref{sec:2}, we briefly review the $f(R)$ gravity in Palatini formalism and study the cosmological dynamical behavior of Palatini $f(R)$ theories. In Section \ref{sec:3}, we apply the statefinder diagnostic to a series of $f(R)$ gravity models. In Section \ref{sec:4}, we illustrate the results obtained from the observational constraints and apply statefinder diagnosis to the best-fits. Finally, the conclusions and the discussions are presented in Section \ref{sec:5}.

\section{The Palatini $f(R)$ gravity and its cosmological dynamics}
\label{sec:2}
\subsection{A brief overview of $f(R)$ gravity in Palatini formalism}
We firstly review the Palatini formalism from the generalized Einstein-Hilbert action
\begin{equation}\label{eq:1}
  S = \frac{1}{2\kappa}\int d^4 x \sqrt{-g} f(R) + S_m(g_{\mu\nu}, \psi),
\end{equation}
where $\kappa \equiv 8\pi G$, $G$ is the gravitational constant, $g$ is the determinant of the metric $g_{\mu\nu}$ (Greek indices such as $\mu$, $\nu$ run through 0...3 throughout the paper), $f(R)$ is the general function of the generalized Ricci scalar $R\equiv g^{\mu\nu}R_{\mu\nu}(\Gamma^{\lambda}_{\mu\nu})$, and $S_m$ is the matter action which depends only upon the metric $g_{\mu\nu}$ and the matter fields $\psi$ and not upon the independent connection $\Gamma^{\lambda}_{\mu\nu}$ that is differentiated from the Levi-Civita connection $\{^{\lambda}_{\mu\nu}\}$. It should be noted that when $f(R) = R$, GR will come about.

Varying the action with respect to the metric $g_{\mu\nu}$ and the connection $\Gamma^{\lambda}_{\mu\nu}$ respectively yields
\begin{equation}\label{eq:2}
  f'(R)R_{\mu\nu}-\frac{1}{2}f(R)g_{\mu\nu}=\kappa T_{\mu\nu},
\end{equation}
and
\begin{equation}\label{eq:3}
  \overline{\nabla}_{\lambda}(\sqrt{-g}f'(R)g^{\mu\nu})=0,
\end{equation}
where $f'(R) \equiv df /dR$, $\overline{\nabla}_{\lambda}$ denotes the covariant derivative associated with the independent connection $\Gamma^{\lambda}_{\mu\nu}$, and $T_{\mu\nu}$ is the energy-momentum tensor given by
\begin{equation}\label{eq:4}
  T_{\mu\nu}=\frac{-2}{\sqrt{-g}}\frac{\delta S_m}{\delta g^{\mu\nu}}.
\end{equation}
If we consider a perfect fluid case, then $T^{\mu\nu}=(\rho+p)u^{\mu}u^{\nu}+pg^{\mu\nu}$, where $\rho$, $p$ and $u^{\mu}$ denote the energy density, the pressure, and the four-velocity of the fluid, respectively. Note that $T \equiv g^{\mu\nu}T_{\mu\nu}=-\rho +3p$.
Based on Eq. \ref{eq:3}, we can define a metric conformal to $g_{\mu\nu}$ as
\begin{equation}\label{eq:5}
  h_{\mu\nu}\equiv f'(R)g_{\mu\nu}.
\end{equation}
Then, we can deduce the connection $\Gamma^{\lambda}_{\mu\nu}$ in terms of the conformal metric $h_{\mu\nu}$
\begin{equation}\label{eq:6}
  \Gamma^{\lambda}_{\mu\nu}=\frac{1}{2}h^{\lambda\sigma}(\partial_{\mu} h_{\nu\sigma}+\partial_{\nu} h_{\mu\sigma}-\partial_{\sigma} h_{\mu\nu}),
\end{equation}
or, equivalently, in regard to $g_{\mu\nu}$
\begin{equation}\label{eq:7}
\begin{aligned}
  \Gamma^{\lambda}_{\mu\nu}=&\frac{1}{2f'(R)}g^{\lambda\sigma}[(\partial_{\mu} (f'(R)g_{\nu\sigma})+\partial_{\nu} (f'(R)g_{\mu\sigma})\\
  &-\partial_{\sigma} (f'(R)g_{\mu\nu})].
\end{aligned}
\end{equation}
The corresponding Ricci tensor under the conformal transformation reads
\begin{equation}\label{eq:8}
  R_{\mu\nu}=\partial_{\lambda}\Gamma^{\lambda}_{\mu\nu}-\partial_{\nu}\Gamma^{\lambda}_{\mu\lambda}+\Gamma^{\lambda}_{\lambda\sigma}\Gamma^{\sigma}_{\mu\nu}-\Gamma^{\lambda}_{\mu\sigma}\Gamma^{\sigma}_{\lambda\nu},
\end{equation}
which can be presented by the Ricci tensor $R_{\mu\nu}(g)$ associated with $g_{\mu\nu}$ as
\begin{equation}\label{eq:9}
\begin{aligned}
  R_{\mu\nu}=&R_{\mu\nu}(g)+\frac{3}{2(f'(R))^2}(\nabla_{\mu}f'(R))(\nabla_{\nu}f'(R))\\
  &-\frac{1}{f'(R)}(\nabla_{\mu}\nabla_{\nu}+\frac{1}{2}g_{\mu\nu}\nabla_{\sigma}\nabla^{\sigma})f'(R),
\end{aligned}
\end{equation}
where $\nabla_{\mu}$ is the covariant derivative with respect to the Levi-Civita connection $\{^{\lambda}_{\mu\nu}\}$ of the metric $g_{\mu\nu}$.
\subsection{FRW cosmology of the Palatini $f(R)$ gravity and numerical results}
Since measurements of Cosmic Microwave Background (CMB) suggest that our universe is spatially flat \citep{Halverson2002,Netterfield2002} at late times, we start with a flat FRW universe with metric
\begin{equation}\label{eq:10}
  ds^2 = -dt^2+a^2(t)(dx^2+dy^2+dz^2),
\end{equation}
where $a(t)$ and $t$ are the scale factor and the cosmic time, respectively.

According to Eqs. \ref{eq:2} and \ref{eq:9}, the modified Friedmann equation can be derived as
\begin{equation}\label{eq:11}
  (H+\frac{\dot{f'}(R)}{2f'(R)})^2=\frac{\kappa (\rho +3p)+f(R)}{6f'(R)},
\end{equation}
where $H\equiv\dot{a}/a$ is the Hubble parameter, and the dot denotes the differentiation with respect to the cosmic time $t$.
In addition, taking the trace of Eq. \ref{eq:2} gives
\begin{equation}\label{eq:12}
  f'(R)R-2f(R)=\kappa T.
\end{equation}
If we assume that the universe only contains dust-like (pressureless) matter at late times, then $T = -\rho_m$, where $\rho_m$ represents the energy density of matter. Furthermore, combining Eq. \ref{eq:12} with the energy conservation equation of matter $ \dot{\rho}_m+ 3H\rho_m=0$, we can express $\dot{R}$ as
\begin{equation}\label{eq:13}
  \dot{R}=-\frac{3H(f'(R)R-2f(R))}{f''(R)R-f'(R)},
\end{equation}
where $f''(R)\equiv d^2f/dR^2$. Substituting Eq. \ref{eq:13} into Eq. \ref{eq:11}, we can obtain
\begin{equation}\label{eq:14}
  H^2=\frac{1}{6f'}\frac{3f-f'R}{[1-\frac{3f''(f'R-2f)}{2f'(f''R-f')}]^2}.
\end{equation}
Since the redshift $z$ can be expressed through $1+z=a_0/a$ and conventionally the present scale factor $a_0=1$ is chosen (subscript 0 denotes the present time value throughout the paper), we can get the expressions $\rho_m=\rho_{m0}(1+z)^3$ and $dz/dt=-H(1+z)$. Therefore, Eqs. \ref{eq:12} and \ref{eq:13} can be rewritten as
\begin{equation}\label{eq:15}
  f'(R)R-2f(R)=-3H_0^2\Omega_{m0}(1+z)^3,
\end{equation}
and
\begin{equation}\label{eq:16}
  \frac{dR}{dz}=-\frac{9H_0^2\Omega_{m0}(1+z)^2}{f''(R)R-f'(R)},
\end{equation}
where $\Omega_{m0}\equiv \kappa \rho_{m0}/3H_0^2$. As a result, Eq. \ref{eq:14} can also be expressed as
\begin{equation}\label{eq:17}
  \frac{H^2}{H_0^2}=\frac{1}{6f'}\frac{3\Omega_{m0}(1+z)^3+f/H_0^2}{[1+\frac{9H_0^2\Omega_{m0}(1+z)^3f'')}{2f'(f''R-f')}]^2}.
\end{equation}

In order to study the cosmological evolution by using Eqs. \ref{eq:15}-\ref{eq:17}, it is prerequisite to obtain the initial conditions: $(R_0, H_0, \Omega_{m0})$. With Eqs. \ref{eq:15} and \ref{eq:17}, choosing units so that $H_0 = 1$ \citep{Amarzguioui2006}, once the explicit expression of $f(R)$ is given, one can solve for $R_0$ with fixed value of $\Omega_{m0}$. On the other hand, in order to understand the cosmological evolution behavior, it is useful to define the effective equation of state
\begin{equation}\label{eq:18}
  w_{\rm eff} = -1+\frac{2}{3}(1+z)\frac{H'}{H},
\end{equation}
where $H'\equiv dH/dz$. Since the deceleration parameter $q$ is related as follows
\begin{equation}\label{eq:19}
  q = -1+(1+z)\frac{H'}{H} = \frac{1}{2}(1+3w_{\rm eff}),
\end{equation}
one can certainly explore the cosmological dynamics through the evolutions of $w_{\rm eff}$ with different models of Palatini $f(R)$ gravity.
\subsubsection{$f(R)$ theories with power-law terms}
We consider the following general form for $f(R)$
\begin{equation}\label{eq:20}
  f(R) = R + \alpha R^m - \beta R^{-n},
\end{equation}
where $m$ and $n$ are real constants with the same sign. Such theories have been investigated to explain the early and the late accelerated expansion of our universe \citep{Sotiriou2006,Meng2004}. Note that not all combinations of $m$ and $n$ are in agreement with a flat universe with the early matter dominated era followed by an accelerated expansion at late times. At the early times of matter-dominated era, the universe is better described by GR in order to avoid confliction with early-time physics such as Big Bang Nucleosynthesis (BBN) and CMB. It implies that the modified Lagrangian should recover the standard GR Lagrangian for large $R$, and hence it requires $m<1$ and $n>-1$.
Then, we consider two specific types of theories in this regime.

\paragraph{\textbf{(i) The $\alpha=0$ case}}~{}

\begin{figure*}
  \centering
  \includegraphics[width=5.8in,height=2.4in]{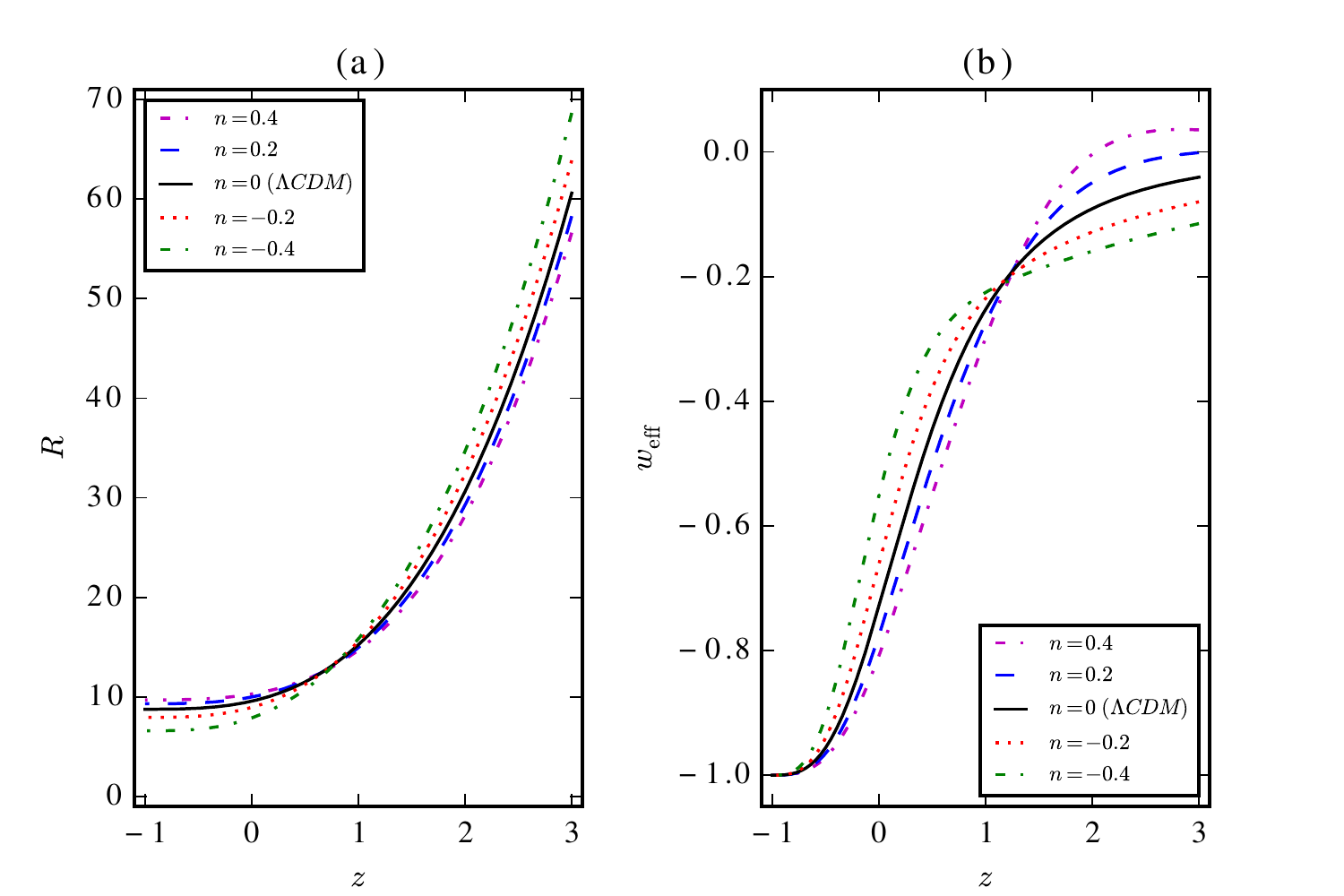}
  \caption{The evolutions of the scalar curvature $R$ and the effective equation of state $w_{\rm eff}$ against redshift $z$ for $f(R) = R - \beta R^{-n}$. Different values of $n$ are chosen, along with $H_0=1$ and $\Omega_{m0}=0.27$.}\label{fig:1}
\end{figure*}

In this case, the form of $f(R)$ reads
\begin{equation}\label{eq:21}
  f(R) = R - \beta R^{-n}.
\end{equation}
Based on Eqs. \ref{eq:16}-\ref{eq:18}, by adopting $H_0=1$ and $\Omega_{m0}=0.27$, we can obtain the evolutions of the scalar curvature $R$ and the effective equation of state $w_{\rm eff}$ that are shown in Fig. \ref{fig:1}. Note that the special case of $(\beta,n)=(4.38,0)$ corresponds to the $\Lambda$CDM model. From Fig. \ref{fig:1}, one can easily see that for any choice of $n$ ($n>-1$), the curvature $R$ and the effective equation of state $w_{\rm eff}$ decrease with the evolution of the universe. Moreover, the smaller $n$ is, the faster $R$ decreases, and the larger the present values of $w_{\rm eff}$ are. Also, the expansion of the universe can turn from a decelerated phase into an accelerated phase, and the universe approaches de Sitter phase in the future. Interestingly, one can detect from Fig. \ref{fig:1} that there is a intermediate convergence zone where the redshift are around 1.2, and the evolutionary trajectories are upsidedown through the convergence zone. From the prospective of mathematics, the convergence zone must be caused by the slope change of the effective equation of state $w_{\rm eff}$, since they all have the same constraints but for different expressions of $f(R)$. From Eq. \ref{eq:20}, one can certainly see that the value of $f(R)$ is sensitive as the power $n$ varies, including its derivatives with respect to $R$. The variances embodied in Fig. \ref{fig:1} are clear: As the redshift grows, at first, the slopes of the curves are slowly increased, and then they become steady, and at last, the slopes slowly decrease to zero in the future. Therefore, the different dropping rates of $w_{\rm eff}$ lead to the convergence. Also, it is worth noticing that in the $n=0.4$ case, the slope undergos from negative to positive.

\paragraph{\textbf{(ii) The $m=n=-1/2$ case}}~{}

\begin{figure*}
  \centering
  \includegraphics[width=5.8in,height=2.4in]{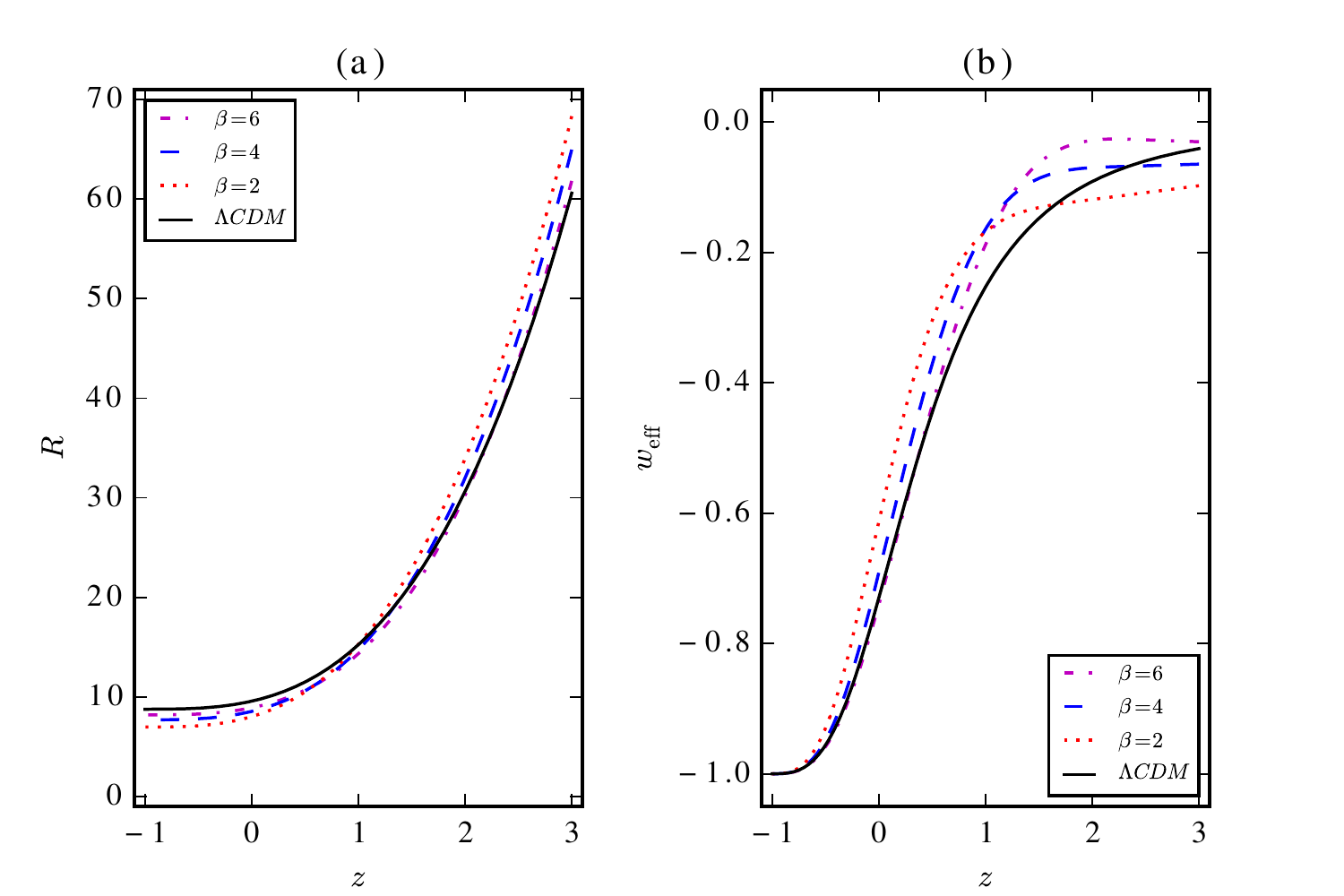}
  \caption{Same as Fig. \ref{fig:1}, except for theories of type $f(R) = R + \alpha R^{1/2} - \beta R^{-1/2}$. Different values of $\beta$ are chosen, along with $H_0=1$ and $\Omega_{m0}=0.27$.}\label{fig:2}
\end{figure*}

In this case, Eq. \ref{eq:20} becomes
\begin{equation}\label{eq:22}
  f(R) = R + \alpha R^{1/2} - \beta R^{-1/2}.
\end{equation}
Fig. \ref{fig:2} is plotted by going through the same procedure as above case. One can ascertain that the curvature $R$ and the effective equation of state $w_{\rm eff}$ decrease with evolution for different choices of $\beta$. Furthermore, the smaller $\beta$ results in the faster decrease of $R$, and larger $w_{\rm eff}$ at the present time. It is also obvious that the universe evolves from deceleration to acceleration, and enters to de Sitter acceleration in the future.

\subsubsection{$f(R)$ theories with logarithm term}

\begin{figure*}
  \centering
  \includegraphics[width=5.8in,height=2.4in]{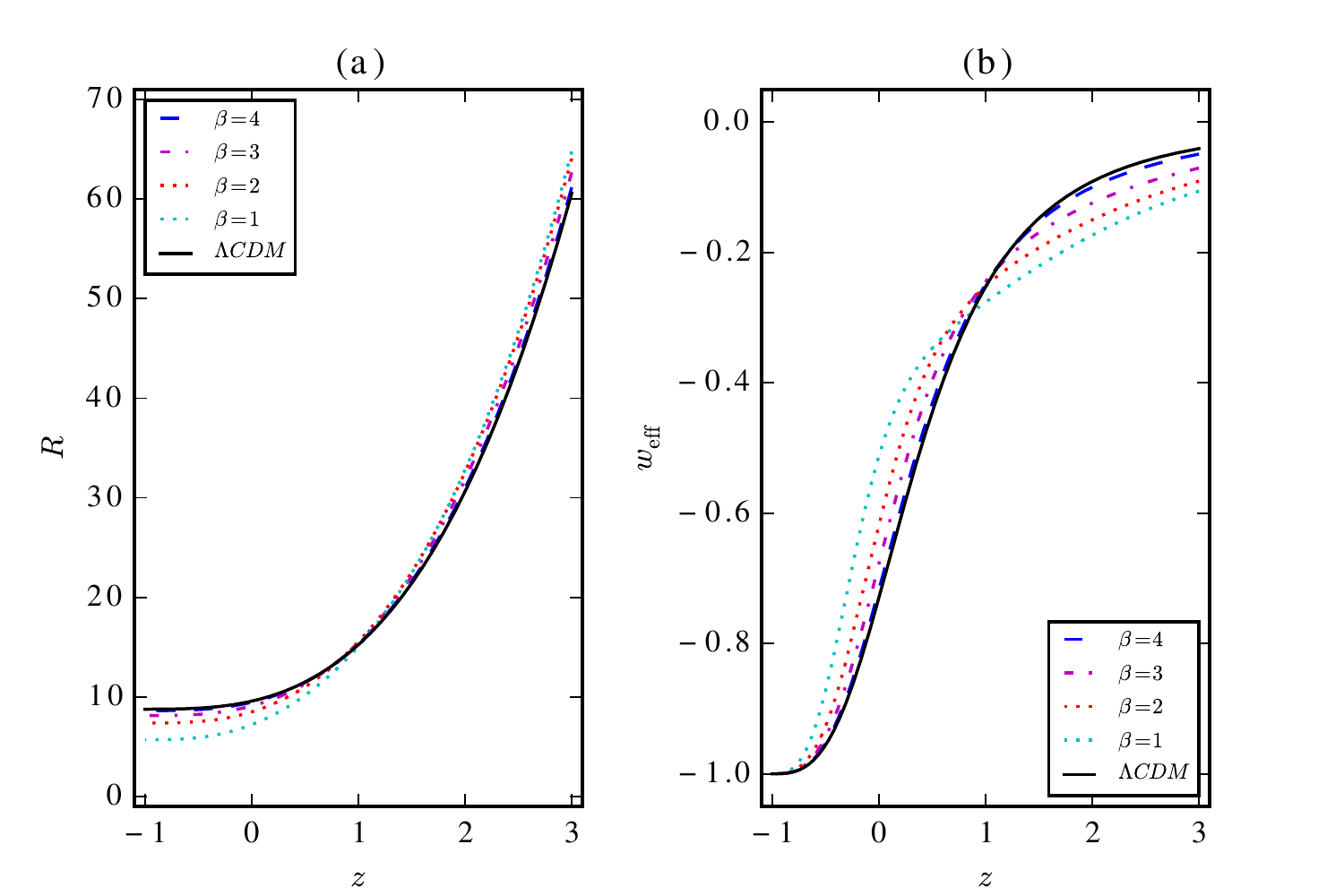}
  \caption{Same as Fig. \ref{fig:1}, except for theories of type $f(R) = R + \alpha \ln R - \beta$.}\label{fig:3}
\end{figure*}

Finally we exploit $f(R)$ theories of the type
\begin{equation}\label{eq:23}
  f(R) = R + \alpha \ln R - \beta,
\end{equation}
which has been studied in \citep{Nojiri2004,Meng2004}, and it has been claimed that such theories have a well-defined Newtonian limit \citep{Nojiri2004}. Note that, the asymptotic behavior lim$_{R\rightarrow \infty}¡Þ f(R)\rightarrow R$ is obtained for any choice of $\alpha$ and $\beta$, and thus, the arbitrary $\alpha$ and $\beta$ can satisfy the assumption that the universe can described by GR at the early times. However, not all combinations of $\alpha$ and $\beta$ can explain a late-time accelerated expansion of the universe. Therefore, for the sake of compatibility with the observational constraints obtained in Sec. \ref{sec:4}, we select a series of values of $\beta$, which can well present the evolvement of the universe from an early-time deceleration to a late-time acceleration (see also Fig. \ref{fig:3}).

Substituting Eq. \ref{eq:23} into Eqs. \ref{eq:16}-\ref{eq:18}, with $H_0=1$ and $\Omega_{m0}=0.27$, the evolvements of the curvature $R$ and the effective equation of state $w_{\rm eff}$ with respect to the redshift $z$ are illustrated in Fig. \ref{fig:3}. Consequently, $R$ and $w_{\rm eff}$ decrease with the evolution of the universe for the set of $\beta$. Also, the larger $\beta$ gets, the slower decrease of $R$ and the smaller $w_{\rm eff}$ at the present time appear to be. Similarly to the results of the above types of theories, the universe evolves from deceleration to acceleration, and gets close to de Sitter universe in the future.

\section{Statefinder diagnostic for the Palatini $f(R)$ gravity}
\label{sec:3}

In this section, we pay attention to the statefinder diagnosis. As we know, two famous geometrical variables characterizing the expansion history of the universe are the Hubble parameter $H$ presenting the expansion rate
of the universe and the deceleration parameter $q \equiv -a\ddot{a}/\dot{a}^2$ characterizing the rate of acceleration/deceleration of the expanding universe. Obviously, they only depend on the scale factor $a$ and its first and second derivatives in terms of $t$, i.e., $\dot{a}$ and $\ddot{a}$. However, with the enhancing amount of cosmological models and the remarkable increasing in the accuracies of cosmological observational data, these two parameters are no longer sensitive enough for distinguishing different models, which can be revealed from the fact that many cosmological models correspond to the same current value of $q$. As a result, the so-called statefinder diagnosis was introduced in order to discriminate more and more cosmological models involving dark energy. It can be constructed using both the second and third derivatives of the scale factor $a$.

In addition to $H$ and $q$, two new parameters are defined as the statefinder pair $\{r,s\}$
\begin{equation}\label{eq:24}
  r\equiv \frac{\dddot{a}}{aH^3},\quad s\equiv \frac{r-1}{3(q-1/2)}.
\end{equation}
Since different cosmological models exhibit distinct evolutionary trajectories in the $r-s$ plane, the statefinder diagnostic is probably a fine tool to distinguish cosmological models. The remarkable property is that $\{r,s\}=\{1,0\}$ corresponds to the $\Lambda$CDM model. So one can clearly identifies the ``distance'' from a given cosmological model to $\Lambda$CDM model in the $r-s$ plane, such as the quintessence, the phantom, the Chaplygin gas, the holographic dark energy models, the interacting dark energy models, which have been studied in the literatures \citep{Alam2003,Zhang2005,Setare2007,Yi2007}. Especially, the current values of the parameters $s$ and $r$ in these diagrams can provide a considerable way to measure the ``distance'' from a given model to $\Lambda$CDM model.

According to Eq. \ref{eq:19}, the statefinder pair $\{r,s\}$ can be rewritten as
\begin{equation}\label{eq:25}
  r=1-2(1+z)\frac{H'}{H}+(1+z)^2(\frac{H'^2}{H^2}+\frac{H''}{H}),
\end{equation}
\begin{equation}\label{eq:26}
  s=\frac{-2(1+z)H'/H+(1+z)^2(H'^2/H^2+H''/H)}{3[-3/2+(1+z)H'/H]},
\end{equation}
where $H''\equiv d^2H/dz^2$.

\begin{figure*}[htbp]
\centering
  \subfloat[]{%
    \includegraphics[width=5.8in,height=2.4in]{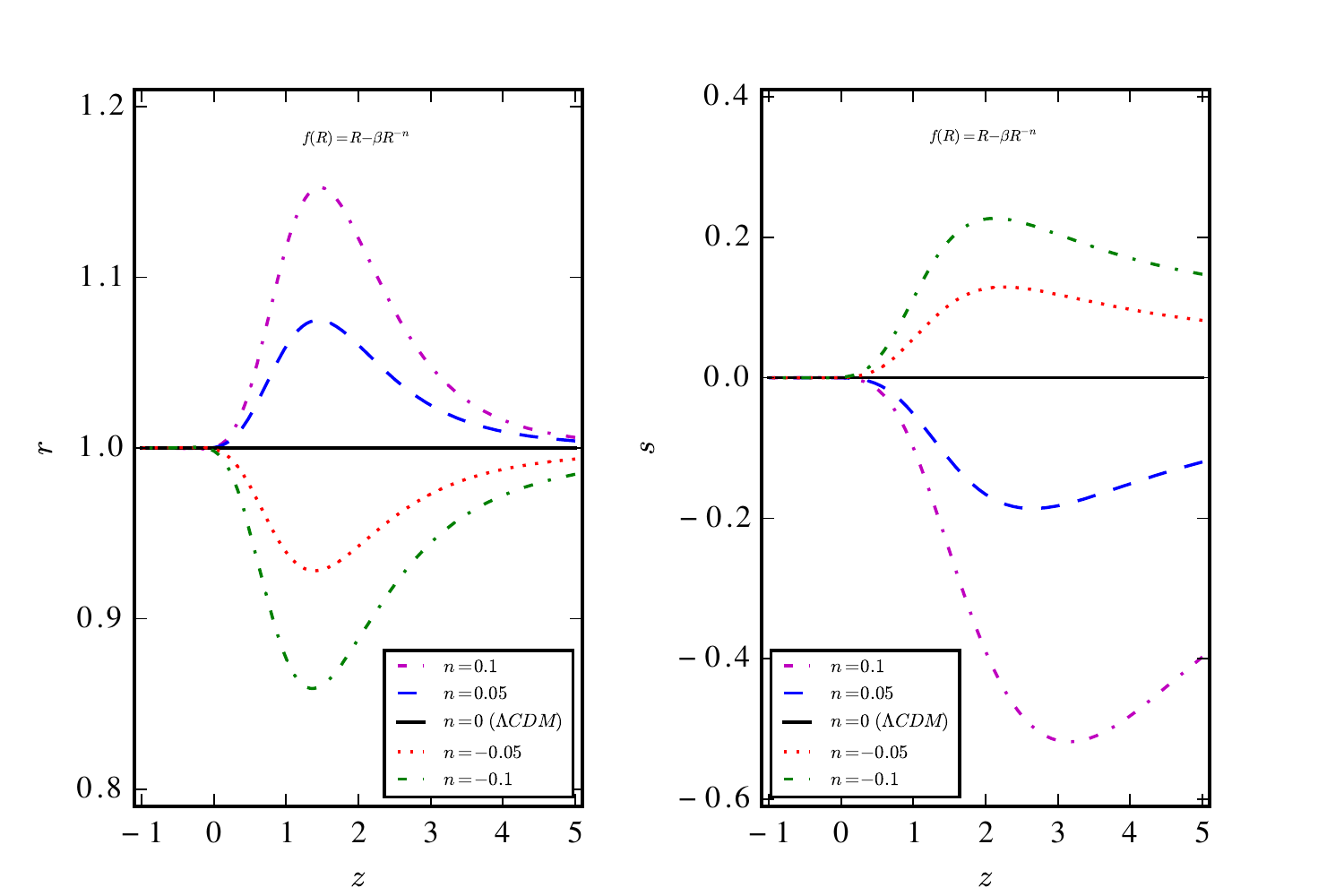}}\\
  \subfloat[]{%
    \includegraphics[width=5.8in,height=2.4in]{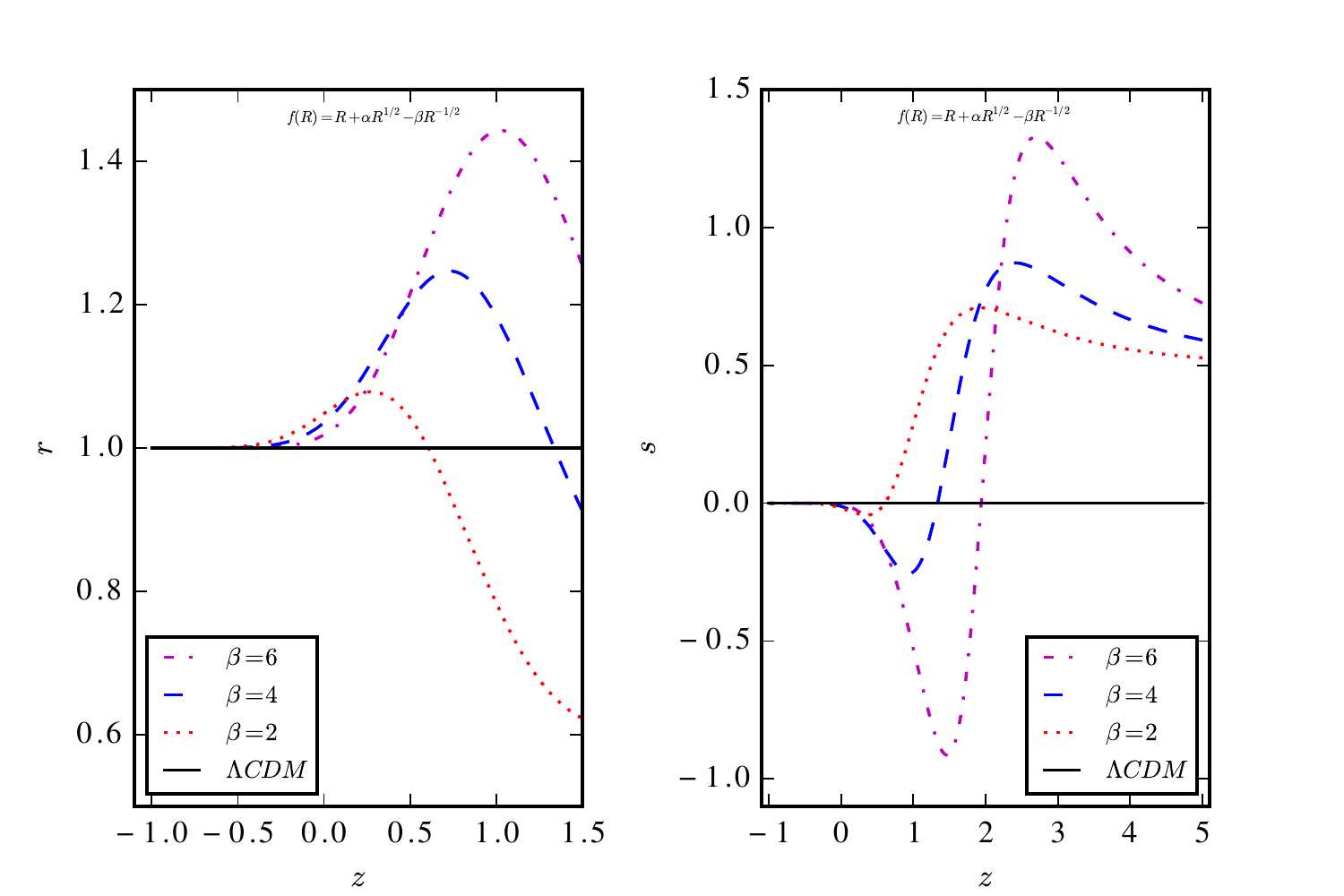}}\\
  \subfloat[]{%
    \includegraphics[width=5.8in,height=2.4in]{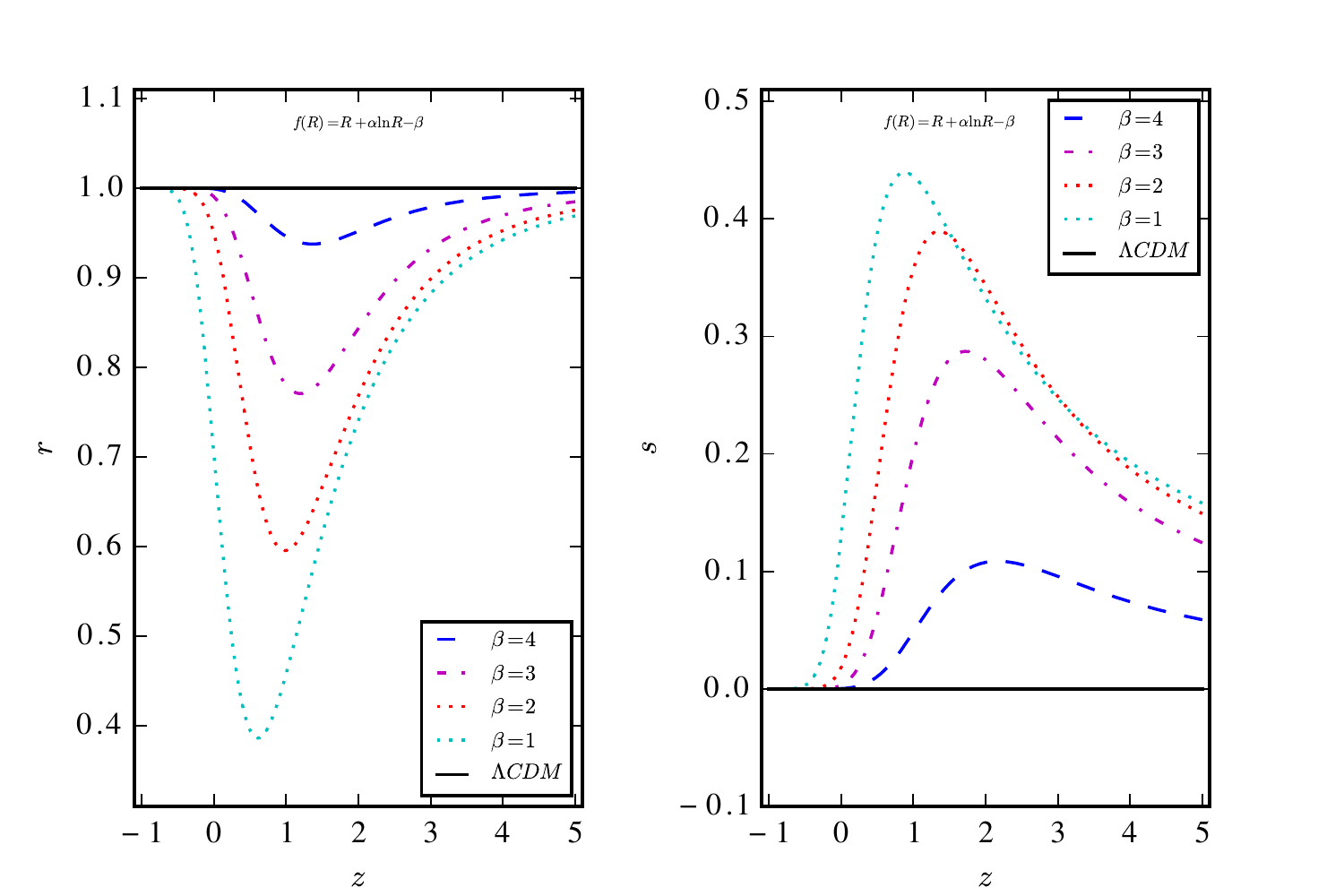}}\\
  \caption{Evolutions of $r(z)$ and $s(z)$ for the theories of type $f(R) = R - \beta R^{-n}$, $f(R) = R + \alpha R^{1/2} - \beta R^{-1/2}$, and $f(R) = R + \alpha \ln R - \beta$ with $H_0=1$ and $\Omega_{m0}=0.27$.}\label{fig:4}
\end{figure*}

\begin{figure*}[htbp]
  \centering
  \includegraphics[width=5.8in,height=2.5in]{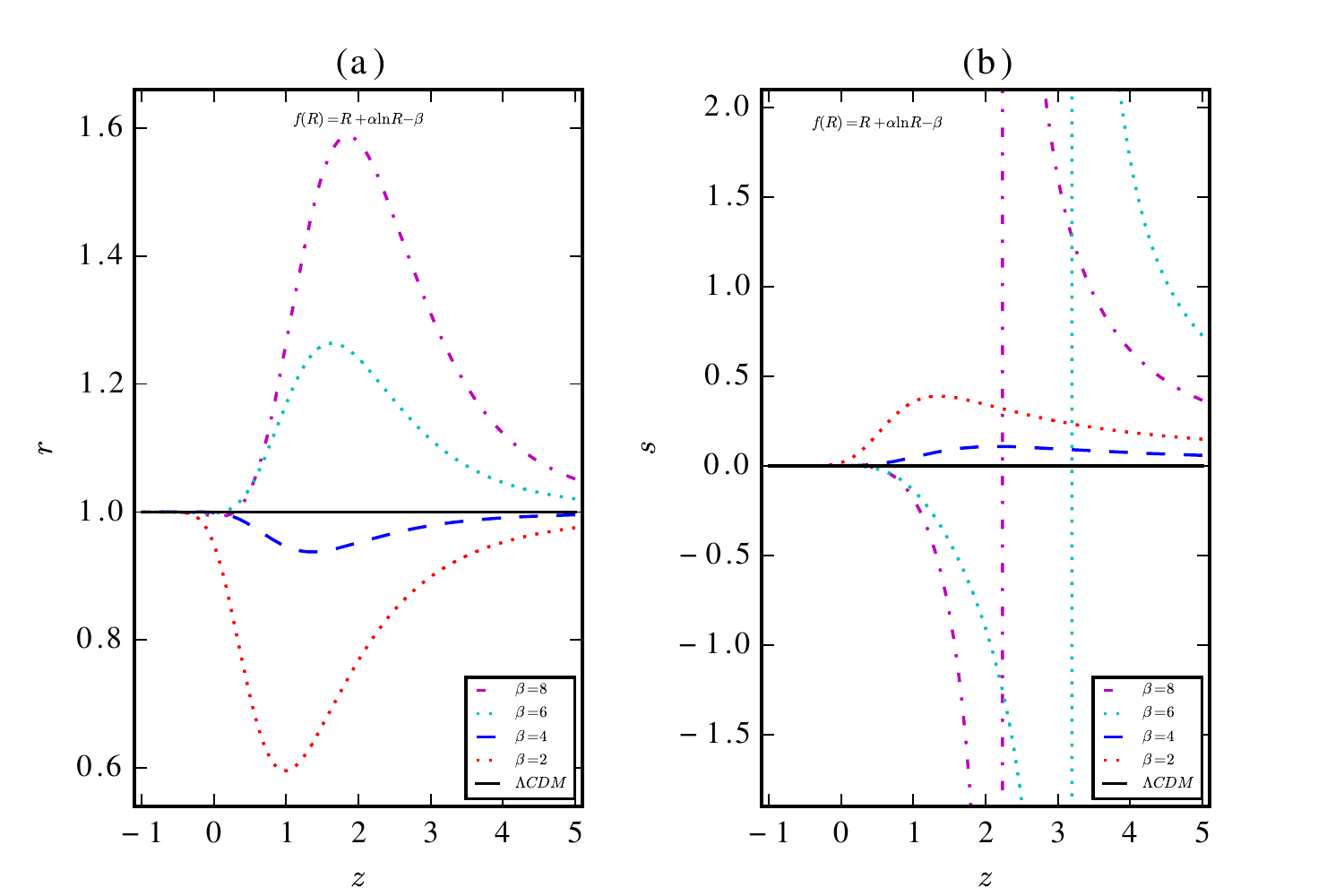}
  \caption{Same as Fig. \ref{fig:4} (c), but for different values of $\beta$. Note that $\beta=6$ and $\beta=8$ cases indicate that at times of $q$ around 0.5, $s$ tends to infinity as expected.}\label{fig:5}
\end{figure*}

In what follows, we employ statefinder diagnostic to the $f(R)$ theories mentioned in Section \ref{sec:2}. However, due to the fact that the singularity comes when the denominator of $s$ tends to zero (i.e., $q=0.5$ case), which can be seen from Fig. \ref{fig:5} (b), the values of parameters we deliberately select in such theories are not all the same as those in the previous section. Comparing Fig. \ref{fig:5} with Fig. \ref{fig:4} (c), one can palpably find that not all combinations of $\alpha$ and $\beta$ are suitable for statefinder diagnosis. Next, we will show that $r-s$ planes display the distinct evolutionary trajectories for these Palatini $f(R)$ theories, and hence one can discriminate various types of Palatini $f(R)$ theories from one another, and not to mention other dark energy models.

\begin{figure*}[htbp]
\centering
  \subfloat[]{%
    \includegraphics[width=5.8in,height=2.4in]{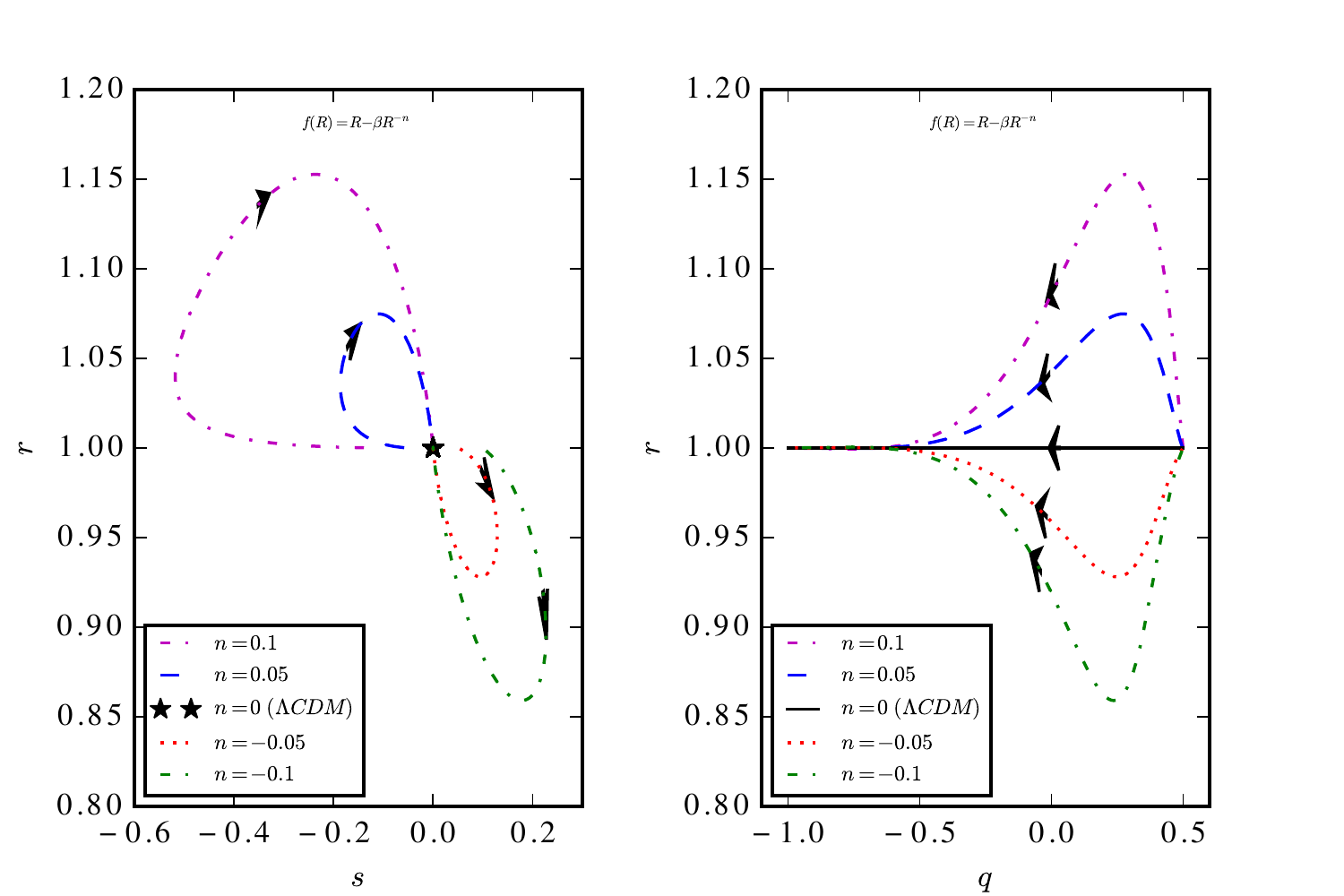}}\\
  \subfloat[]{%
    \includegraphics[width=5.8in,height=2.4in]{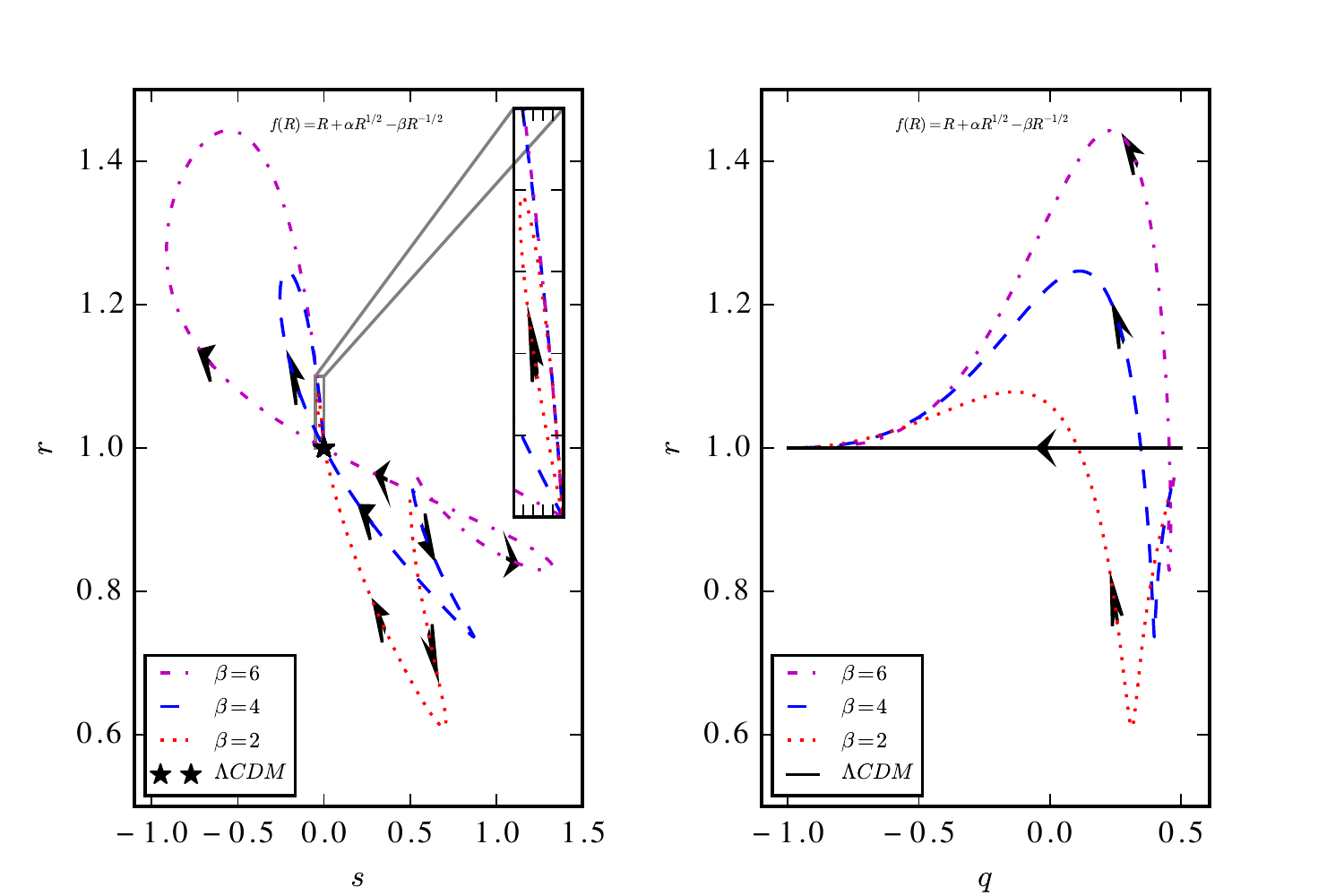}}\\
  \subfloat[]{%
    \includegraphics[width=5.8in,height=2.4in]{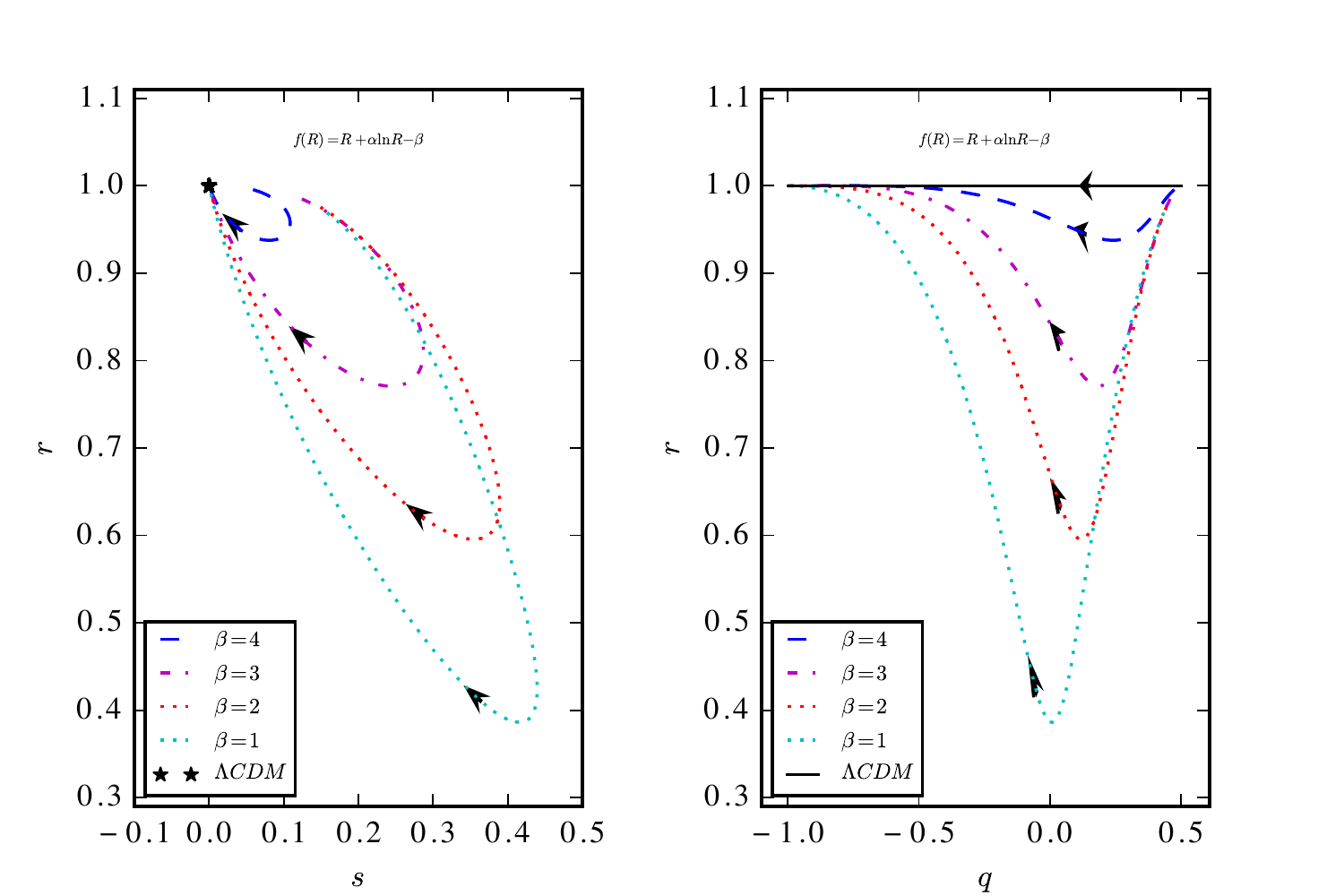}}\\
  \caption{Statefinder diagnostic $r-s$ and $r-q$ planes for the theories of type $f(R) = R - \beta R^{-n}$, $f(R) = R + \alpha R^{1/2} - \beta R^{-1/2}$, and $f(R) = R + \alpha \ln R - \beta$ with $H_0=1$ and $\Omega_{m0}=0.27$.}\label{fig:6}
\end{figure*}

Fig. \ref{fig:4} demonstrates the evolutions of $r$ and $s$ with respect to redshift $z$ for $f(R)$ theories mentioned above. It can be shown that different features are exhibited as follows:
\begin{enumerate}
  \item For the model $f(R) = R - \beta R^{-n}$ (see also Fig. \ref{fig:4} (a)), the curves stay at one side of the $\Lambda$CDM line ($r = 1$ and $s = 0$). Specifically, for $n>0$ case, the evolutionary curves lie in the region of $r>1, s<0$, and for $n<0$ case, inversely, they remain at region $r<1, s>0$. Moreover, the trajectories of evolution are all first moving away from the $\Lambda$CDM line and then towards it. This could evidently be revealed from $r-s$ planes as shown in Fig. \ref{fig:6} (a). In addition, the larger the absolute values of $n$ become, the further the traces of evolvement move from the $\Lambda$CDM line. Eventually, they both tends to evolve like $\Lambda$CDM universe (de-Sitter point, i.e., \{r,s\}=\{1,0\}, or \{q,r\}=\{-1,1\}) in the future.
  \item As for the theories of type $f(R) = R + \alpha R^{1/2} - \beta R^{-1/2}$ in Fig. \ref{fig:4} (b), we can easily find that for any choices of $\beta$, the evolutionary curves cross the $\Lambda$CDM line sooner or later, and the larger $\beta$ gets, the bigger the fluctuations of $r$ and $s$ turn. Also, they all will come to an end like $\Lambda$CDM universe in the future.
  \item The models of type $f(R) = R + \alpha \ln R - \beta$ explored in Fig. \ref{fig:4} are all the cases that $\alpha$ and $\beta$ have the opposite sign (note that only for the $\beta>0$ case), and in comparison with Fig. \ref{fig:5} (a) where $\alpha,\beta>0$ holds, one can realize that evolutions of $r$ depend on the sign of $\alpha$ and $\beta$. In the former case, $r$ lies in the region $r<1$, and also $s>0$. It is worth mentioning that the ``distance'' between the evolutionary trajectories and the $\Lambda$CDM line grows smaller along with larger values of $\beta$. In the latter case, inversely, $r$ lies in the region $r>1$, and the larger $\beta$ are, the farther $r$ moves away from $r=1$ line. However, both cases turn towards $\Lambda$CDM case in the future.
\end{enumerate}

Anyhow, it is evidently seen from above features that the trajectories of evolutions varies from different choices of parameters and from model to model. Finally, $r-s$ and $r-q$ planes are plotted in Fig. \ref{fig:6}. Observing $r-s$ plane is more clear than just looking at the separate evolutions of $r$ and $s$, especially when it comes to compare between cosmological models. After all, it is more palpable for $r-s$ planes with distinct evolutionary trajectories and explicit evolutionary directions to tell the differences. $r-s$ and $r-q$ planes significantly exhibit the deviations between the Palatini $f(R)$ theories, and also show that deceleration/acceleration transition occurs in these models. Therefore, the statefinder diagnostic is a fair way to differentiate various cosmological models.

\section{Constraints with OHD}
\label{sec:4}
In order to determine if these $f(R)$ models are compatible with cosmological observations, here we intend to constrain the parameters in above types of $f(R)$ models with OHD. The current available OHD dataset is listed in table \ref{tab:Hubble}. With this dataset, we can adopt these $f(R)$ theories to Eq. \ref{eq:17}, and for the goodness of fit we employ the standard $\chi ^2$ minimization, defined by

\begin{equation}\label{eq:27}
  \chi ^2=\sum_i\frac{[H_{\rm th}(z_i|\textbf{p})-H_{\rm obs}(z_i)]^2}{\sigma^2(z_i)},
\end{equation}
where $H_{\rm th}(z_i|\textbf{p})$ is the theoretical Hubble parameter at redshift $z_i$ given by Eq. \ref{eq:17}, and $\textbf{p}$ depends on the $f(R)$ models; $H_{\rm obs}(z_i)$ are the OHD, and $\sigma(z_i)$ is the uncertainty of each $H_{\rm obs}(z_i)$. Note that the covariance matrix of data is not necessarily diagonal, as discussed in \cite{Yu2013}, and if not, the case will become complicated and should be treated by means of the method mentioned by \cite{Yu2013}. Here we assume that each measurement in $\{H_{\rm obs}(z_i)\}$ is independent.

In what follows we will proceed to constrain on the models studied in the previous section. When calculating $\chi ^2$, we first exploit $H_0=1$ to Eqs. \ref{eq:15}-\ref{eq:17}, and then make the resulting $H_{\rm th}(z_i)$ to be multiplied by iterated values of $H_0$. Subsequently, we marginalize the parameters to plot contour figures. Meanwhile, with the best-fits of each model, we apply statefinder diagnostic to them.

\begin{table}
\centering
\begin{tabular}{|lccc|}
\hline
{$z$}   & $H(z)$ & Method & Ref.\\
\hline
$0.0708$   &  $69.0\pm19.68$      &  I    &  \cite{Zhang2014}   \\
    $0.09$       &  $69.0\pm12.0$        &  I    &  \cite{Jimenez2003}   \\
    $0.12$       &  $68.6\pm26.2$        &  I    &  \cite{Zhang2014}  \\
    $0.17$       &  $83.0\pm8.0$          &  I    &  \cite{Simon2005}     \\
    $0.179$     &  $75.0\pm4.0$          &  I    &  \cite{Moresco2012}     \\
    $0.199$     &  $75.0\pm5.0$          &  I    &  \cite{Moresco2012}     \\
    $0.20$         &  $72.9\pm29.6$        &  I    &  \cite{Zhang2014}   \\
    $0.240$     &  $79.69\pm2.65$      &  II   &  \cite{Gaztanaga2009}   \\
    $0.27$       &  $77.0\pm14.0$        &  I    &    \cite{Simon2005}   \\
    $0.28$       &  $88.8\pm36.6$        &  I    &  \cite{Zhang2014}   \\
    $0.35$       &  $84.4\pm7.0$          &  II   &   \cite{Xu2013}  \\
    $0.352$     &  $83.0\pm14.0$        &  I    &  \cite{Moresco2012}   \\
    $0.3802$     &  $83.0\pm13.5$        &  I    &  \cite{Moresco2016}   \\
    $0.4$         &  $95\pm17.0$           &  I    &  \cite{Simon2005}     \\
    $0.4004$     &  $77.0\pm10.2$        &  I    &  \cite{Moresco2016}   \\
    $0.4247$     &  $87.1\pm11.2$        &  I    &  \cite{Moresco2016}   \\
    $0.43$     &  $86.45\pm3.68$        &  II   &  \cite{Gaztanaga2009}   \\
    $0.44$       & $82.6\pm7.8$           &  II   &  \cite{Blake2012}  \\
    $0.4497$     &  $92.8\pm12.9$        &  I    &  \cite{Moresco2016}   \\
    $0.4783$     &  $80.9\pm9.0$        &  I    &  \cite{Moresco2016}   \\
    $0.48$       &  $97.0\pm62.0$        &  I    &  \cite{Stern2010}     \\
    $0.57$       &  $92.4\pm4.5$          &  II   &  \cite{Samushia2013}   \\
    $0.593$     &  $104.0\pm13.0$      &  I    &  \cite{Moresco2012}   \\
    $0.6$         &  $87.9\pm6.1$          &  II   &  \cite{Blake2012}   \\
    $0.68$       &  $92.0\pm8.0$          &  I    &  \cite{Moresco2012}   \\
    $0.73$       &  $97.3\pm7.0$          &  II   &  \cite{Blake2012}  \\
    $0.781$     &  $105.0\pm12.0$      &  I    &  \cite{Moresco2012}   \\
    $0.875$     &  $125.0\pm17.0$      &  I    &  \cite{Moresco2012}   \\
    $0.88$       &  $90.0\pm40.0$        &  I    &  \cite{Stern2010}     \\
    $0.9$         &  $117.0\pm23.0$      &  I    &  \cite{Simon2005}  \\
    $1.037$     &  $154.0\pm20.0$      &  I    &  \cite{Moresco2012}   \\
    $1.3$         &  $168.0\pm17.0$      &  I    &  \cite{Simon2005}     \\
    $1.363$     &  $160.0\pm33.6$      &  I    &  \cite{Moresco2015}  \\
    $1.43$       &  $177.0\pm18.0$      &  I    &  \cite{Simon2005}     \\
    $1.53$       &  $140.0\pm14.0$      &  I    &  \cite{Simon2005}     \\
    $1.75$       &  $202.0\pm40.0$      &  I    &  \cite{Simon2005}     \\
    $1.965$     &  $186.5\pm50.4$      &  I    &   \cite{Moresco2015}  \\
    $2.34$       &  $222.0\pm7.0$        &  II   &  \cite{Delubac2015}   \\
\hline
\end{tabular}
\caption{\label{tab:Hubble} The current available OHD dataset. The method I is the differential galactic ages method, and II represents the radial BAO method. $H(z)$ is in units of ${\rm km/s/Mpc}$ here.}
\end{table}

\subsection{Theories of the type $f(R) = R - \beta R^{-n}$}

\begin{figure*}[htbp]
  \centering
  \includegraphics[width=5.8in,height=3.5in]{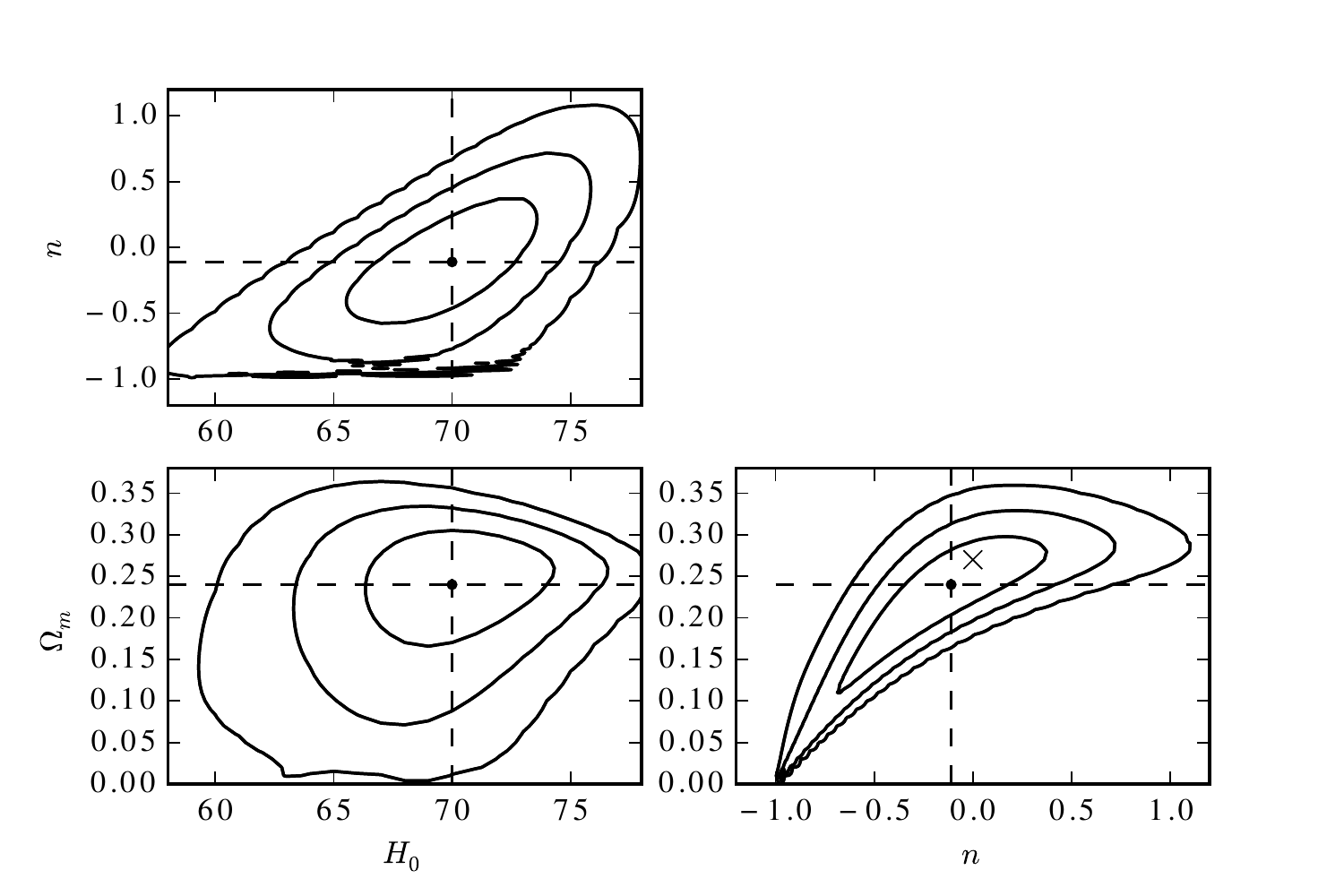}
  \caption{1$\sigma$, 2$\sigma$, and 3$\sigma$ confidence regions of the theory $f(R) = R - \beta R^{-n}$ constrained by OHD. The dashed line and the dot present the best-fit values of $H_0=70\ \rm km/s/Mpc$, $\Omega_{m0}=0.24$, and $n=-0.11$. The cross mark represents the $\Lambda$CDM model.}\label{fig:7}
\end{figure*}

In this context, the set of parameters is selected with $\textbf{p}=(H_0,\Omega_{m0},n)$. Fig. \ref{fig:7} shows the constraints in the 1$\sigma$, 2$\sigma$, and 3$\sigma$ confidence regions, in which contour plots of the two out of three parameters are presented by marginalizing the third parameter. The best-fit values of $\textbf{p}$ are $H_0=70\ \rm km/s/Mpc$, $\Omega_{m0}=0.24$, and $n=-0.11$, along with the corresponding value of $\beta=3.65$. In the combined analysis of \cite{Amarzguioui2006}, the best-fit model is found to be $\beta=3.6$ and $n=-0.09$, which is consistent with our best-fits. After marginalizing over $H_0$, $\Omega_{m0}$ and $n$, the 1$\sigma$ constraint values are ($\Omega_{m0}=0.24^{+0.06}_{-0.13}$, $n=-0.11^{+0.49}_{-0.58}$), ($H_0=70^{+3.6}_{-4.5}\ \rm km/s/Mpc$, $n=-0.09^{+0.47}_{-0.60}$), and ($\Omega_{m0}=0.25^{+0.055}_{-0.084}$, $H_0=70^{+4.3}_{-3.7}\ \rm km/s/Mpc$), respectively. Note that $\Lambda$CDM model lies in the 1$\sigma$ confidence level, which corresponds to $\Omega_{m0}=0.27$, and $n=0$, marked by cross in Fig. \ref{fig:7}.

Since we have the best-fit model of this type, the statefinder diagnostic can be exploited to study its real evolutionary process. As shown in Fig. \ref{fig:10}, the evolutionary trajectories are indeed compatible with the features described in previous section for $n<0$ case (see also Fig. \ref{fig:6} (a)). Evidently, one can find the best-fit model of this type is capable of producing late-time acceleration of the universe and includes the deceleration/acceleration transition stage.

\subsection{Theories of the type $f(R) = R + \alpha R^{1/2} - \beta R^{-1/2}$}

\begin{figure*}
  \centering
  \includegraphics[width=5.8in,height=3.5in]{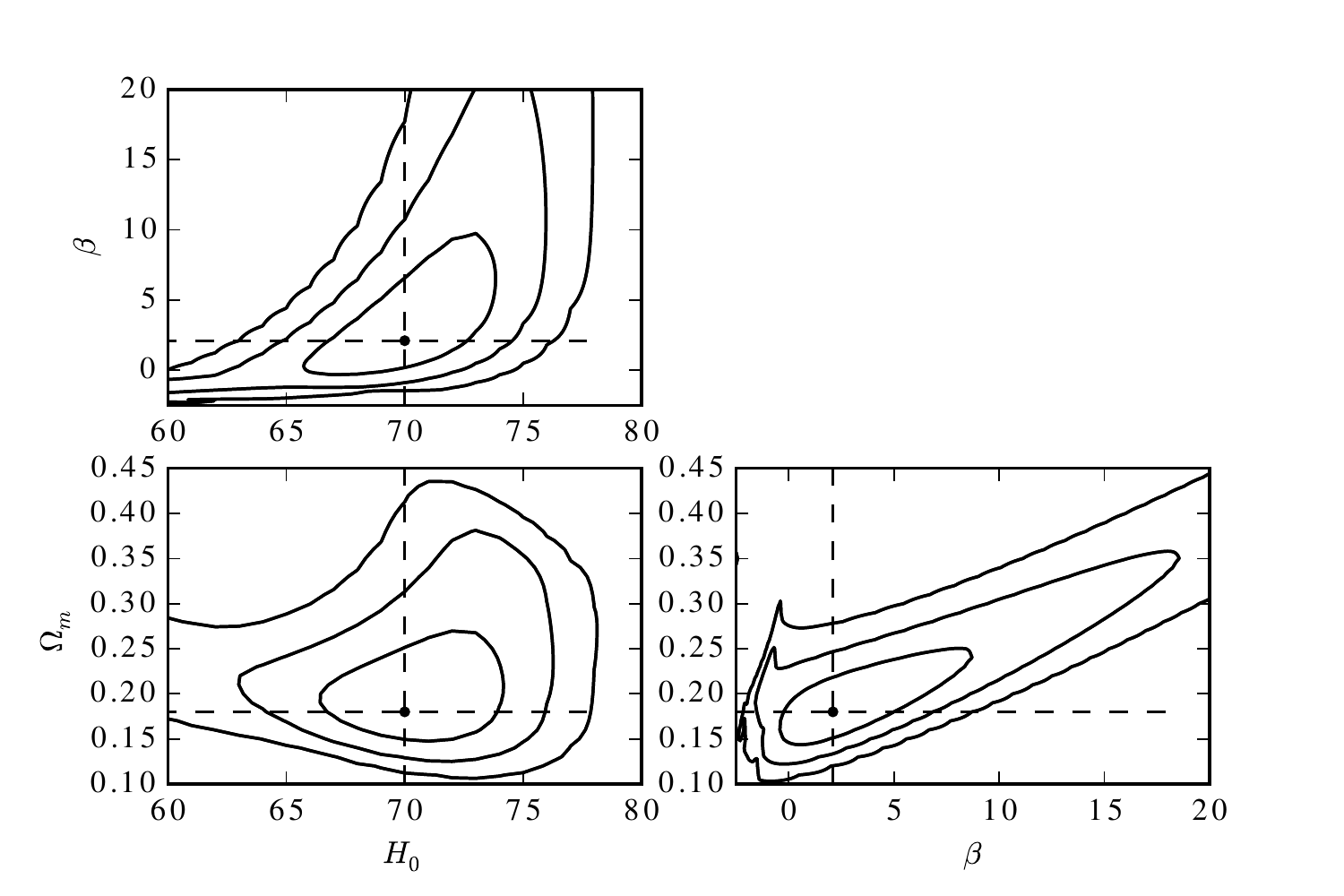}
  \caption{Same as Fig. \ref{fig:7}, but for the theory $f(R) = R + \alpha R^{1/2} - \beta R^{-1/2}$. The dashed line and the dot show the best-fit values of $H_0=70\ \rm km/s/Mpc$, $\Omega_{m0}=0.18$, and $\beta=2.1$.}\label{fig:8}
\end{figure*}

In this circumstance, by choosing $\textbf{p}=(H_0,\Omega_{m0},\beta)$, we plot the contour figures with the same methods as previous model in Fig. \ref{fig:8}. The best-fits of $\textbf{p}$ are $H_0=70\ \rm km/s/Mpc$, $\Omega_{m0}=0.18$, and $\beta=2.1$ associated with $\alpha=-1.53$. By marginalizing over $H_0$, $\Omega_{m0}$ and $\beta$ separately, we obtain the corresponding 1$\sigma$ constraints of ($\Omega_{m0}=0.18^{+0.073}_{-0.039}$, $\beta=2.2^{+6.60}_{-2.72}$), ($H_0=70^{+3.96}_{-4.34}\ \rm km/s/Mpc$, $\beta=2.6^{+7.23}_{-3.12}$), and ($\Omega_{m0}=0.19^{+0.082}_{-0.045}$, $H_0=71^{+3.28}_{-4.64}\ \rm km/s/Mpc$), respectively. This shows a example where the two nonlinear terms of Eq. \ref{eq:20} are comparable and necessary to produce the acceleration at late times as \cite{Fay2007} stated. However, our best-fits varies from them in the $m=n=1/2$ case, which may be caused by the different set of parameters being constrained.

In Fig. \ref{fig:10}, $r-s$ and $r-q$ planes indicate that they fit the characteristics explored in section \ref{sec:3}, and the trajectories of evolvement clearly differ from the other two types of $f(R)$ theories in Palatini formalism. Also, the best-fit model of this type is able to evolve from decelerated expansion to the late-time acceleration of the universe.

\subsection{Theories of the type $f(R) = R + \alpha \ln R - \beta$}

\begin{figure*}
  \centering
  \includegraphics[width=5.8in,height=3.5in]{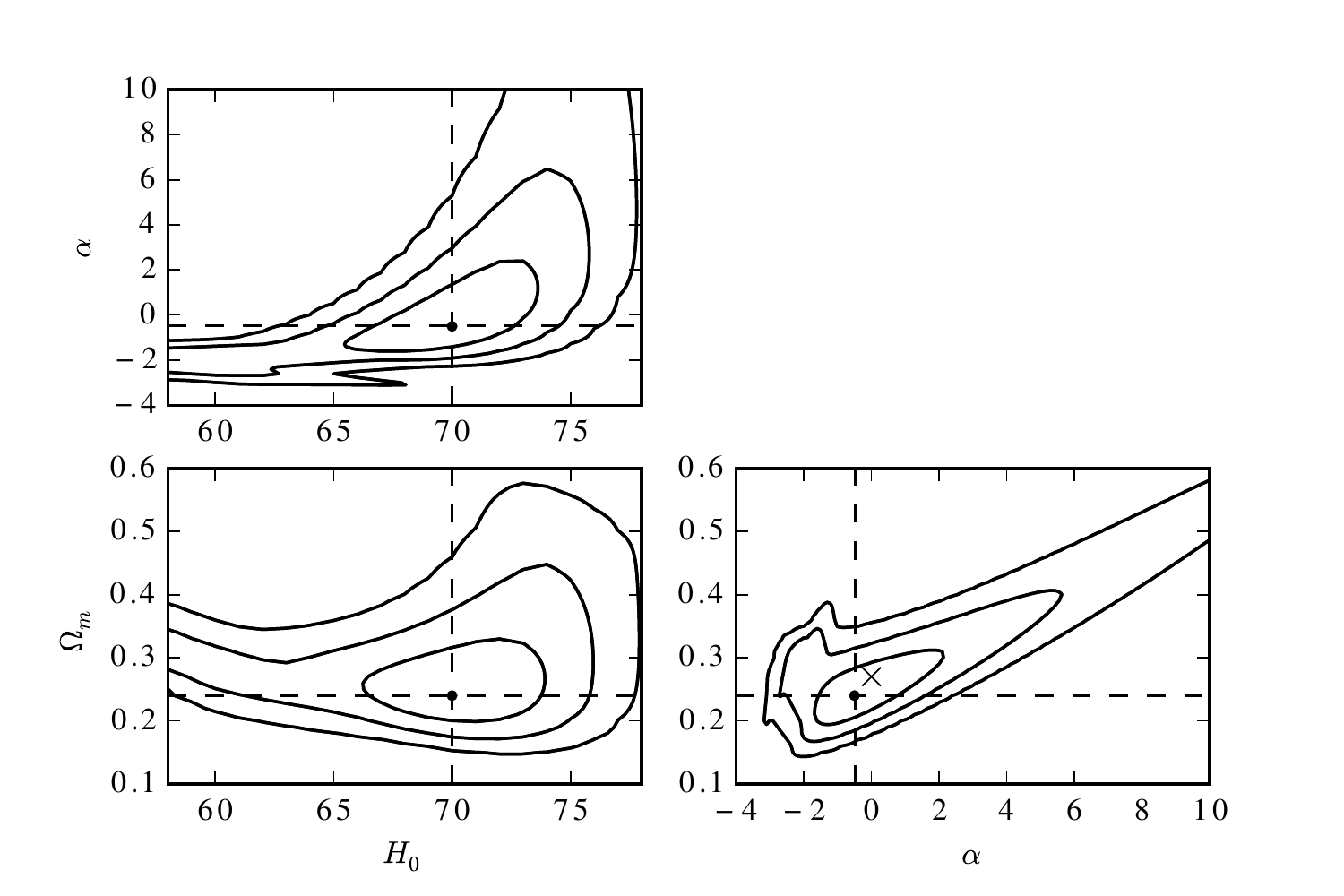}
  \caption{Same as Fig. \ref{fig:7}, but for the theory $f(R) = R + \alpha \ln R - \beta$. The dashed line and the dot exhibit the best-fit values of $H_0=70\ \rm km/s/Mpc$, $\Omega_{m0}=0.24$, and $\alpha=-0.48$. The cross mark represents the $\Lambda$CDM model.}\label{fig:9}
\end{figure*}

In this situation, we set $\textbf{p}$ to be $(H_0,\Omega_{m0},\alpha)$. Same as above methods, the confidence regions are demonstrated in Fig. \ref{fig:9}, where the best-fit constraints are $H_0=70\ \rm km/s/Mpc$, $\Omega_{m0}=0.24$, and $\alpha=-0.48$ coupled with $\beta=3.58$. Consequently, they are not compatible with the best-fit model of \cite{Fay2007} which corresponds to $\alpha=0.11$ and $\beta=4.62$, and excludes $\beta=0$ case. While our constraints include $\beta=0$ case (when $\alpha$ around -2.3) in the 2$\sigma$ region, it means that the assertion made by \cite{Fay2007} that the $\ln R$ term alone cannot drive the late-time acceleration without cosmological constant, is not tenable, let alone the fact that $\Lambda$CDM model ($\alpha=0$ and $\Omega_{m0}=0.27$, marked in Fig. \ref{fig:9}) is well contained in the 1$\sigma$ region. The marginalized 1$\sigma$ constraints are ($\Omega_{m0}=0.24^{+0.076}_{-0.050}$, $\alpha=-0.48^{+2.67}_{-1.26}$), ($H_0=70^{+3.7}_{-4.6}\ \rm km/s/Mpc$, $\alpha=-0.3^{+2.80}_{-1.44}$), and ($\Omega_{m0}=0.25^{+0.082}_{-0.055}$, $H_0=71^{+4.0}_{-3.8}\ \rm km/s/Mpc$).

\begin{figure*}
  \centering
  \includegraphics[width=5.8in,height=3in]{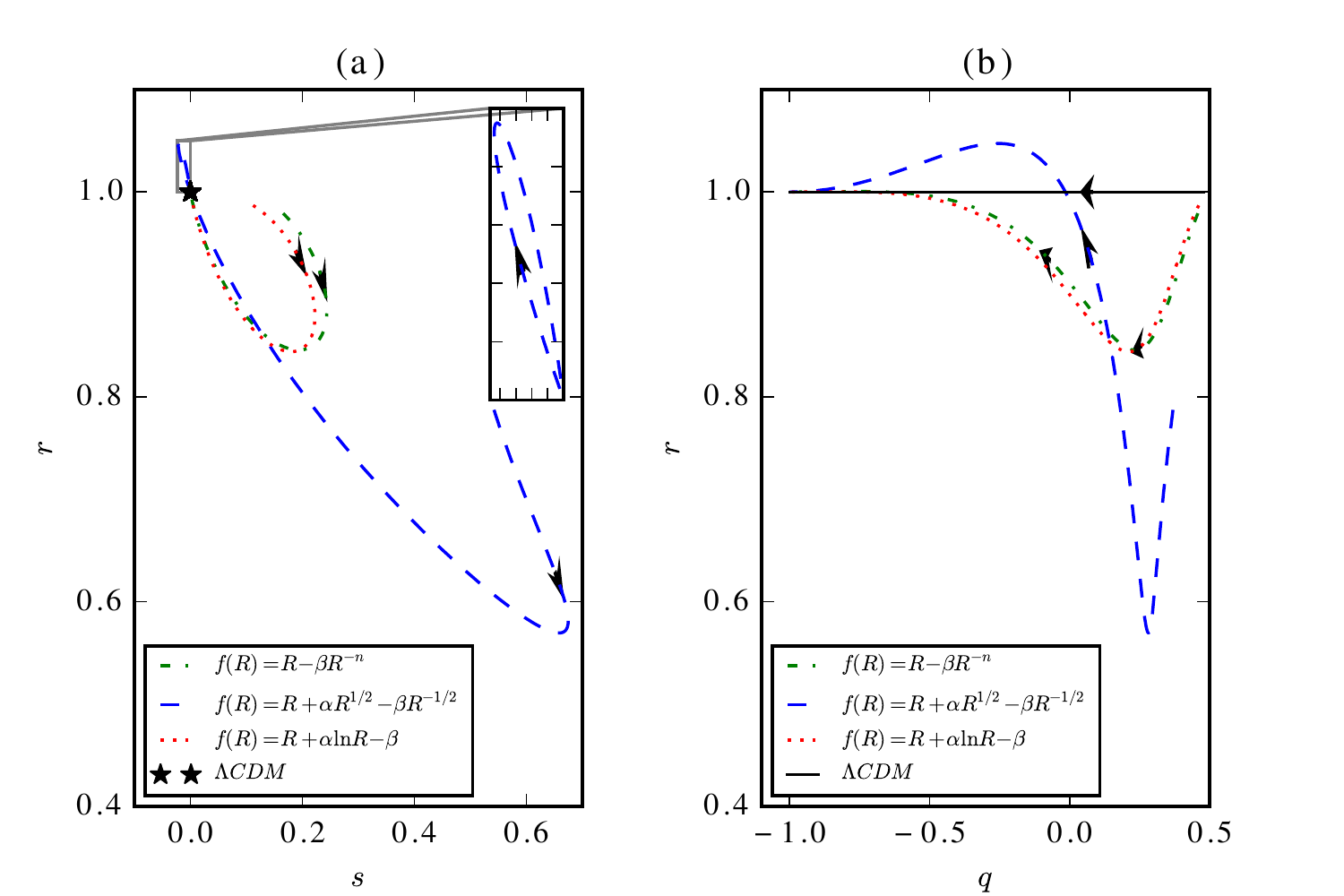}
  \caption{Statefinder pairs $r-s$ and $r-q$ planes for the best-fit models of the Palatini $f(R)$ theories of type $f(R) = R - \beta R^{-n}$, $f(R) = R + \alpha R^{1/2} - \beta R^{-1/2}$, and $f(R) = R + \alpha \ln R - \beta$.}\label{fig:10}
\end{figure*}

As shown in Fig. \ref{fig:10}, as a common feature of the type $f(R) = R - \beta R^{-n}$ and the type $f(R) = R + \alpha R^{1/2} - \beta R^{-1/2}$ that we missed in section \ref{sec:3}, the evolutionary trajectories of the best-fit models are almost indistinguishable in $r-q$ plane, but still distinct in $r-s$ plane. Therefore, this characteristic is further in favour of the statefinder pair $\{r,s\}$ in the direction of discriminating different cosmological models. Also, $r-q$ plane shows that the best-fit model of this type can explain the transition of phase from decelerated expansion to the late-time acceleration of the universe.

\section{Conclusions and discussions}
\label{sec:5}
In previous sections, we have systematically studied the cosmological dynamics of a series of types of $f(R)$ theories within Palatini approach and applied the statefinder diagnostic to these models, and then placed observational constraints on the parameters of the models.

First, we find that different features of evolutionary trails of the Ricci curvature $R$ and the effective equation of state $w_{\rm eff}$ with respect to redshift $z$ are revealed. For all the models of Palatini $f(R)$ theory, the values of the order index $n$ ($n>-1$) of $R$ for Eq. \ref{eq:21}, the parameter $\beta$ of Eq. \ref{eq:22} and the parameter $\beta$ of Eq. \ref{eq:23} all have negative correlations with the decreased rate of $R$, and positive correlations with the fluctuations of $w_{\rm eff}$.

Second, since more and more theories proposed to account for the late-time accelerated expansion of the universe, the well-known parameters, such as the Hubble parameter $H$, the deceleration parameter $q$, and equation of state $w$, are not enough to discriminate these models. Especially, for the case of modified gravity theories such as string/M-theory, extended scalar-tensor models, and braneworld models of dark energy, the equation of state $w$ is not a fundamental physical entity. The more general and sensitive diagnosis known as statfinder diagnostic emerges as required. However, it makes one wonder if this diagnosis can stand the trial at all times. Therefore we employ it to the $f(R)$ theories of types Eqs. \ref{eq:21}, \ref{eq:22}, and \ref{eq:23}, to see if it can still hold well for the sake of discrimination. Eventually, one can draw conclusions that the trajectories of evolutions vary from model to model and with different set of values for the given parameters, and also $r-s$ and $r-q$ planes further exhibit clarity of differences among the models. As a result, the $f(R)$ theories with chosen series of parameters not only display distinct trajectories of evolvement, but also present evident deceleration/acceleration transition along with late-time acceleration and tendency towards $\Lambda$CDM model in the future. Thus, the statefinder diagnostic holds for this literature as a efficient way to distinguish between cosmological models. We believe that further explorations for its validity will be made in future researches for more and more cosmological models, and even if it somewhat fails, more advanced diagnosis would be proposed. In some sense, now that $\dot{a}$, $\ddot{a}$, and $\dddot{a}$ are involved, the fourth even fifth order derivatives of the scale factor $a$ are more probable to be included to enhance the accuracy of the diagnostic.

Third, we exploit OHD to obtain observational constraints on the there models Eqs. \ref{eq:21}, \ref{eq:22}, and \ref{eq:23}. On the one hand, the best-fits of Eq. \ref{eq:21} are compatible with the combined constraints of \cite{Amarzguioui2006}, but not consistent with the constraints of \cite{Fay2007}. Also, the goodness of fit is apart from \cite{Fay2007} for both models of Eqs. \ref{eq:22} and \ref{eq:23}. On the other hand, for the first model Eq. \ref{eq:21}, the $\Lambda$CDM model lies in the 1$\sigma$ confidence region, as well as the type Eq. \ref{eq:23}. The seconde model Eq. \ref{eq:22} shows a example where the two nonlinear terms of (20) are comparable and necessary to produce the acceleration at late times. As for the third model Eq. \ref{eq:23}, the constraints include $\beta=0$ case in the 2$\sigma$ confidence region, which means that the $\ln R$ term alone can possibly drive the late-time acceleration without cosmological constant.

Ultimately, we employ the statefinder diagnostic to the three best-fit models. In consequence, for one thing, we find that the evolutionary trajectories have the same properties described in section \ref{sec:3}. For another thing, it is worth noticing that the common features between best-fit models of Eqs. \ref{eq:22} and \ref{eq:23}. On the $r-q$ plane, one can almost not discriminate their trails, but along with $r-s$ plane as a complement, obviously, the differences between them are certain. This further admit the merits of the statefinder pair $\{r,s\}$. In addition, all the best-fit models of the Palatini $f(R)$ theory have the same properties as follows: (i) They both carry out the earlier-time deceleration and the late-time acceleration phase in the matter-dominated universe; (ii) They both tend to turn into $\Lambda$CDM cosmology in the future. Notice that in the paper we only consider the flat late-time matter-dominated FRW universe, which is perfectly suitable for OHD for the low redshift range of $0.0708<z<2.34$.

\begin{acknowledgements}
This work was supported by the National Science Foundation of China (Grants No. 11573006, 11528306, 11347163), the Fundamental Research Funds for the Central Universities£¬and the Special Program for Applied Research on Super Computation of the NSFC-Guangdong Joint Fund (the second phase), the Science and Technology Program Foundation of the Beijing Municipal Commission of Education of China under Grant No. KM201410028003.
\end{acknowledgements}

\appendix
\section{Relevant derivatives}
In what follows, we give some derivatives in terms of the generalized Ricci curvature $R$, and the redshift $z$, which can facilitate the statefinder diagnostic to plot easier.

According to Eq. \ref{eq:15}, we define parameter $A$ as follows
\begin{equation}\label{eq:28}
\begin{aligned}
    A&\equiv[1-\frac{3f''(f'R-2f)}{2f'(f''r-f')}]^2\\
     &=[1+\frac{9f''\Omega_{m0}H_0^2(1+z)^3}{2f'(f''r-f')}]^2,
\end{aligned}
\end{equation}
and then the Hubble parameter $H$ becomes
\begin{equation}\label{eq:29}
\begin{aligned}
  H=\sqrt{\frac{3f-f'R}{6f'A}}=\sqrt{\frac{3\Omega_{m0}H_0^2(1+z)^3+f}{6f'A}}.
\end{aligned}
\end{equation}
The first derivative of $A$ in terms of $R$ reads
\begin{equation}\label{eq:30}
\begin{aligned}
  A'\equiv\frac{dA}{dR}=&-3\sqrt{A}\{\frac{f'R-2f}{f''R-f'}[\frac{f'''}{f'}-\frac{f''^2}{f'^2}\\
&-\frac{f''f'''R}{f'(f''R-f')}]+\frac{f''}{f'}\}.
\end{aligned}
\end{equation}
The second derivative of $A$ is written as
\begin{equation}\label{eq:31}
\begin{aligned}
  A''\equiv&\frac{d^2A}{dR^2}=\frac{A'^2}{2A}-3\sqrt{A}\{(\frac{f''''}{f'}-\frac{f'R-2f}{f''R-f'}\frac{3f''f'''}{f'^2}\\
&+\frac{2f''^3}{f'^3})+2(\frac{f'''}{f'}-\frac{f''^2}{f'^2})[1-\frac{f'''R(f'R-2f)}{(f''R-f')^2}]\\
&+\frac{f''}{f'(f''R-f')}[\frac{2f'''^2R^2(f'R-2f)}{(f''R-f')^2}\\
&-\frac{(f'''+f''''R)(f'R-2f)}{f''R-f'}]-f'''R\}.
\end{aligned}
\end{equation}
The first derivative of $R$ with respect to redshift $z$ is shown in Eq. \ref{eq:16}, and the first derivative of $H$ in terms of $R$ is given as
\begin{equation}\label{eq:32}
  \frac{dH}{dR}=\frac{1}{2HA}[\frac{1}{3}+\frac{A'}{2A}(\frac{R}{3}-\frac{f}{f'})-\frac{ff''}{2f'^2}].
\end{equation}
Thus $H'$ relates to $dH/dR$ as follows
\begin{equation}\label{eq:33}
  H'=\frac{dH}{dR}\frac{dR}{dz}=-\frac{dH}{dR}\frac{9\Omega_{m0}H_0^2(1+z)^2}{f''R-f'}.
\end{equation}
The second derivative of $H$ with respect $R$ is expressed as
\begin{equation}\label{eq:34}
\begin{aligned}
  \frac{d^2H}{dR^2}=&-\frac{(dH/dR)^2}{H}+\frac{1}{2HA}\{\frac{A'}{A}[\frac{A'}{A}(\frac{f}{f'}-\frac{R}{3})\\
&+\frac{ff''}{f'^2}-\frac{2}{3}]-\frac{1}{2}[\frac{f''}{f'}+\frac{f}{f'}\\
&+\frac{f}{f'}(\frac{A''}{A}+\frac{f'''}{f'})]+\frac{ff''^2}{f'^3}+\frac{A''R}{6A}\},
\end{aligned}
\end{equation}
which corresponds to $H''$ as follows
\begin{equation}\label{eq:35}
  H''=\frac{dH}{dR}\frac{d^2R}{dz^2}+\frac{d^2H}{dR^2}(\frac{dR}{dz})^2,
\end{equation}
where the second derivative of $R$ relating to $z$ is
\begin{equation}\label{eq:36}
  \frac{d^2R}{dz^2}=\frac{dR}{dz}(\frac{2}{1+z}-\frac{f'''R}{f''R-f'}\frac{dR}{dz}),
\end{equation}

\bibliographystyle{raa}
\bibliography{mybibfile}

\end{document}